\documentclass[prd,amsmath,amssymb,superscriptaddress,nofootinbib]{revtex4-2}
\usepackage{graphicx,url,amssymb,amsmath,rotating,color,units,wasysym,epsfig,multirow,epstopdf,subcaption,xcolor}
\graphicspath{{figures/}}
\usepackage[normalem]{ulem}
\usepackage{comment}
\usepackage[multiple]{footmisc}
\usepackage[colorlinks,urlcolor=blue,citecolor=blue,linkcolor=blue]{hyperref}
\usepackage{cleveref}
\usepackage{multirow}
\usepackage{graphicx}
\usepackage{arydshln}
\usepackage{booktabs}
\usepackage{footnote}

\newcommand{\amplfi}{{\tt AMPLFI}}
\newcommand{\lalsim}{{\tt lalsimulation}}
\newcommand{\aframe}{{\tt Aframe}}
\newcommand{\mbta}{{\tt MBTA}}
\newcommand{\gstlal}{{\tt GstLAL}}
\newcommand{\pycbchyperbank}{{\tt PyCBC-Hyperbank}}
\newcommand{\pycbc}{{\tt PyCBC}}
\newcommand{\spiir}{{\tt SPIIR}}
\newcommand{\mlgw}{{\tt ml4gw}}
\newcommand{\resnet}{{\tt ResNet}}
\newcommand{\resnettwod}{{\tt ResNet2D}}
\newcommand{\resnettf}{{\tt ResNet-34}}
\newcommand{\htcondor}{{\tt HTCondor}}
\newcommand{\tcn}{{\tt TCN}}
\newcommand{\bilby}{{\tt Bilby}}

\newcommand{\aligonoise}{{\tt aLIGOZeroDetHighPower}}

\newcommand{\phenomp}{\tt IMRPhenomPv2}
\newcommand{\phenompnrt}{\tt {IMRPhenomPv2\_NRTidal}}

\newcommand{\mc}{\ensuremath{\mathcal{M}}}
\newcommand{\msun}{\ensuremath{M_{\odot}}}
\newcommand{\mpc}{\ensuremath{\mathrm{Mpc}}}

\newcommand{\boldtheta}{\ensuremath{\mathbf{\Theta}}}

\newcommand{\whiteneddata}{\ensuremath{\tilde{\mathbf{d}}_w}}
\newcommand{\heterodata}{\ensuremath{\mathbf{d^H}}}
\newcommand{\heterodatapooled}{\ensuremath{\mathbf{\bar{d}^H}}}

\DeclareMathOperator*{\argmax}{arg\,max}

\begin{document}
\preprint{LIGO-Pxxxxxx}

%\title{Search and parameter estimation for gravitational waves from low-mass binaries in real-time}

%\title{Machine learning-based search for gravitational waves from binary neutron stars at matched filter sensitivity}

%\title{Machine learning-based search for gravitational waves from binary neutron stars at optimal sensitivity}

%trying to compact it;
%\title{Machine learning for gravitational-waves searches of binary neutron stars at optimal sensitivity}

% And even more????
% \title{AI for gravitational-waves searches of binary neutron stars at optimal sensitivity}

% a slight change of wording
\title{AI-enabled gravitational-waves searches for binary neutron stars at optimal sensitivity}

\author{Bhavya Gupta}
\affiliation{LIGO Laboratory, 185 Albany St, MIT, Cambridge, MA 02139, USA}
\author{Deep Chatterjee}
\affiliation{LIGO Laboratory, 185 Albany St, MIT, Cambridge, MA 02139, USA}
\affiliation{Department of Physics, MIT, Cambridge, MA 02139, USA}
\author{William Benoit}
\affiliation{School of Physics and Astronomy, University of Minnesota, Minneapolis, MN 55455, USA}
\author{Ethan Marx}
\affiliation{LIGO Laboratory, 185 Albany St, MIT, Cambridge, MA 02139, USA}
\affiliation{Department of Physics, MIT, Cambridge, MA 02139, USA}
\author{Christina Reissel}
\affiliation{LIGO Laboratory, 185 Albany St, MIT, Cambridge, MA 02139, USA}
\affiliation{Department of Physics, MIT, Cambridge, MA 02139, USA}
\author{Seiya Tsukamoto}
\affiliation{School of Physics and Astronomy, University of Minnesota, Minneapolis, MN 55455, USA}
\author{Kyungseop Yoon}
\affiliation{LIGO Laboratory, 185 Albany St, MIT, Cambridge, MA 02139, USA}
\affiliation{Department of Physics, MIT, Cambridge, MA 02139, USA}
\author{Michael W. Coughlin}
\affiliation{School of Physics and Astronomy, University of Minnesota, Minneapolis, MN 55455, USA}
\author{Philip Harris}
\affiliation{Department of Physics, MIT, Cambridge, MA 02139, USA}
\author{Erik Katsavounidis}
\affiliation{LIGO Laboratory, 185 Albany St, MIT, Cambridge, MA 02139, USA}
\affiliation{Department of Physics, MIT, Cambridge, MA 02139, USA}

\date{\today}

\begin{abstract}
Gravitational Waves (GWs) represent the newest window of astronomy, furthering our understanding of compact objects like black holes and neutron stars in the Universe. The signal from two merging neutron stars is especially interesting since it brings the prospect 
of concordant electromagnetic and neutrino emissions that can be captured by telescopes on the ground and in space. 
Such multi-messenger observations have a transformational impact on fundamental physics, understanding nuclear matter, astrophysics, and gravity; it was first witnessed in 2017 with the detection of GW170817, a binary neutron star (BNS) merger that established multi-messenger astrophysics with gravitational waves~\cite{gw170817}. 
However, searching for BNS signals in real-time in the ground-based LIGO-Virgo-KAGRA (LVK) GW detectors presents a computational challenge, 
as the data streaming out of the instruments must be matched against $\sim$ million reference waveforms, which requires up to a thousand CPU cores. 
We present a different approach using neural networks to learn the presence of a signal in the data. Our algorithm, called {\aframe}~\cite{aframe-methods},
was deployed in the LVK's fourth observing run and was the first artificial intelligence (AI)-enabled search to detect multiple binary black holes (BBHs) live.
In this work, we demonstrate that the approach 
extends to the lower-mass BNS regime,
and is the first AI-enabled search that achieves sensitivity comparable to matched-filter pipelines at lower computational and latency costs.
The challenge of the longer-duration signals for BNS systems (compared to BBH ones)
is addressed by heterodyning the data, following which the earlier network architectures used for BBHs are sufficient to distinguish segments containing signal versus background. We also show that this analysis is computationally more efficient, requiring a single non-flagship GPU for online deployment. Furthermore, the design and adoption of inference-as-a-service tools allow rapid offline analysis using a distributed pool of GPU resources. Hence, aside from the use case of rapid online data analysis, we also establish the use of {\aframe} for efficient archival data analysis.
\end{abstract}

\maketitle

\section{Introduction}  
\label{sec:intro}
Gravitational-wave (GW) events from compact binary coalescences (CBC) are commonplace today
with one discovery every 2-3 days when data is collected by the LIGO~\cite{advanced_ligo},
Virgo~\cite{advanced_virgo}, and KAGRA~\cite{kagra} GW observatories.
The number of CBCs has grown by two orders of magnitude over the last decade
since the first discovery in 2015~\cite{gw150914}. The most recent LIGO-Virgo-KAGRA's (LVK) fourth observing run,
O4, has seen over 250 candidate events
reported live from online analyses,~\footnote{\url{https://gracedb.ligo.org/superevents/public/O4/}}
while the catalog, GWTC-5~\cite{gwtc5-intro, gwtc5-methods, gwtc5-results, gwtc5-pop, gwtc5-cosmo}, which reports offline %discoveries
search results
up to the second segment of O4 (O4b), lists over 300 events cumulatively since the first detection
of GWs. Beyond the growing event count, the diversity of searches operating on LVK live data has also expanded.
In O4, in addition to the nominal all-sky searches operating in real-time and targeting binary neutron stars (BNS) and stellar-mass
binary black hole (BBH) mergers~\cite{emfollowupguide}, early-warning
template banks~\cite{Sachdev_2020,Nitz_2020,mbta,Kovalam_2022} were deployed live~\cite{emfollowupguide} to identify BNS
mergers prior to coalescence to alert follow-up instruments about potential prompt EM emission. Also, 
a subsolar-mass (SSM) template bank aimed at detecting exotic subsolar-mass binaries was deployed live~\cite{emfollowupguide} for
potential EM follow-up.
%However,
This increase in several live algorithms poses a computational challenge,
as template-based searches typically require $\mathcal{O}(\text{million})$ templates~\cite{gstlal-templates} to efficiently capture
the CBC parameter space and use up to 1000 CPU cores~\cite{computing_resources}. Furthermore, the template bank is expected to grow in size as the
low frequency
detector sensitivity increases in order to preserve the mismatch between templates~\cite{owen-templates, gw-first-det, gstlal-templates}. A different and
promising alternative is the use of function-approximators like neural networks. The literature
on the use of neural networks in search for GWs has grown over the recent
years~\cite{PhysRevD.97.044039,PhysRevLett.120.141103,schafer2020,PhysRevD.105.043002,PhysRevD.107.023021,PhysRevD.103.102003,PhysRevD.106.042002, spiir-ml, huerta2021}.
In particular, O4 saw the first deployment of \aframe~\cite{aframe-methods}, a neural-network-based search, and an
accompanying neural-network based inference pipeline \amplfi~\cite{amplfi}.
Additionally, a machine learning-based method was deployed for the search of unmodeled transients~\cite{MLy}.
A total of 23 event candidates were
detected by {\aframe} in O4~\cite{S250830m,S250830bp,S250901cb,S250904ae,S250904br,
S250904cv,S250911ac,S250927ck,S250927cy,S250929c,S251006dd,S251013x,S251014cn,
S251018bi,S251021u,S251026bn,S251031cq,S251103f,S251105aj,S251108dn,S251108fi,
S251116en,S251117dq}. 
These were sent out as real-time public alerts -- all of these were also
picked up by other template based searches. For a subset of these events, the source properties
from \amplfi\; were sent as part of the alert packets for the astronomy community to ingest
for the purpose of automated follow-up efforts in the electromagnetic spectrum as well as with neutrinos.

\aframe, as deployed in O4, is predominantly sensitive in the binary black hole (BBH) regime reaching
comparable sensitivity to template based searches $\gtrsim 25\;\msun$ at fixed false alarm rate
(see Fig.~4 in ~\cite{aframe-methods}). The longer duration signals for lower mass systems is a challenge
for the approach taken by \aframe. This is because \aframe\; involves a neural-network inference
on a fixed time window (referred to as kernel length subsequently). For the version deployed in O4c, a $1.5$
second kernel length was used. Since {\aframe} performs inference in time-domain, longer signals
associated with lower mass binaries can have a significant portion of the signal content outside the kernel.
Hence deterioration in performance is observed in the search efficiency.
In this work, we address this limitation and develop an extension to {\aframe} suited
to search for low mass binaries like BNS systems. The key idea is to heterodyne, or
demodulate, the data using several simple reference waveforms that compress the power in the signal close
to the merger. Heterodyning is a well known technique in signal processing, and has been used
in GW data analysis for over a decade. In recent years, it has been employed extensively to accelerate parameter estimation and likelihood evaluations~\cite{cornish2013fastfishermatriceslazy,Venumadhav_2019,Cornish_2021,PhysRevD.101.083030,labrador}. More recently, heterodyning-based ratio-filtering approaches have been developed to reduce the effective dimensionality of long-duration waveforms in matched-filter searches for BNS and SSM compact binaries~\cite{pycbc_ratio_filter, ssm_ratio_filter, ratiofilter2026}. \citet{labrador} draws a nice parallel between group equivariance
and the heterodyning transform if done with an approximately correct GW template. Practically, however,
the approximate template is not known apriori for a \emph{search}, the objective of which is to
identify whether a signal is present in the data. If however there is a signal, heterodyning with
a sparse set of GW templates that span the low-mass parameter range, correct up to Newtonian order,
effectively compresses the signal power in regions near the merger. This is because
most of the evolution of a low mass binary involves the inspiral phase. 
Similar data compression techniques
have also been used in addressing parameter estimation for BNSs~\cite{Wong_2023,Dax_2025}.
This is beneficial in two ways: (1) the long duration
signals are compressed from $\mathcal{O}(\text{min.}) \rightarrow \mathcal{O}(\text{sec.})$, and (2)
a neural-network architecture that has seen success with {\aframe} for BBHs, with second-long signals,
is a promising candidate for long duration signals after the heterodyne transform. We show
that this is indeed the case, thus extending the capability of {\aframe} to BNS signals.
We demonstrate the performance using two independent approaches: first using the performance
of GW search pipelines on LVK's GWTC-3~\cite{Abbott_2023}, and second on an independently
generated mock dataset that was used by the LVK for real-time pipeline
benchmarking in O4~\cite{Chaudhary_2024}.

The remainder of this paper is organized as follows. Section~\ref{sec:aframe} describes the high-level overview of the {\aframe} search pipeline, details the heterodyne preprocessing, the change in network architecture input, and the training procedure. Section~\ref{sec:performance} presents the search sensitivity, comparison with matched-filter and existing machine-learning searches, and an evaluation of the model's long-term stability. Section~\ref{sec:mdc_results} summarizes results from the O3 mock data challenge (MDC) dataset. Section~\ref{sec:online_deployment} discusses computational performance and prospects for low-latency deployment. Finally, Section~\ref{sec:conclusion} presents our conclusions and future directions.

\section{The Aframe search}\label{sec:aframe}
{\aframe} is a GW search algorithm that employs neural networks to detect CBCs in GW data collected by the LVK network.
In contrast to traditional template-based algorithms like {\gstlal}~\cite{gstlal}, \texttt{PyCBC}~\cite{pycbc}, {\mbta}~\cite{mbta},
and {\spiir}~\cite{spiir}, that rely on $\mathcal{O}(\mathrm{million})$ bank of waveforms to match against the collected data, {\aframe} 
learns to detect signals by training a neural-network to identify segments of data containing a signal.
{\aframe} was deployed live during the latter half of the LVK's O4 observing run and contributed to the detection
of multiple binary black hole (BBH) candidates. A total of 23 BBH events were reported as public alerts
\cite{S250830m,S250830bp,S250901cb,S250904ae,S250904br, S250904cv,S250911ac,S250927ck,S250927cy,S250929c,
S251006dd,S251013x,S251014cn, S251018bi,S251021u,S251026bn,S251031cq,S251103f,S251105aj,S251108dn,S251108fi,
S251116en,S251117dq} in which {\aframe}, alongside other template-based pipelines, identified these mergers.
In the following subsection, we provide a brief overview of the {\aframe} algorithm and network architecture,
and then describe its extension to lower-mass systems.

\subsection{Search for Binary Black Holes}\label{subsec:bbh}
The {\aframe} neural network is based on a modified {\resnettf}~\cite{resnet} architecture adapted for time-series data.
The model inputs GW strain data from the LIGO Hanford (H1) and Livingston (L1) interferometers as two channels,
and outputs a scalar detection statistic indicating the presence of a signal. The primary change in the architecture,
compared to image-based {\resnet}, is replacing the two-dimensional convolutions with one-dimensional convolutions to operate on timeseries inputs. Additionally, we replace Batch normalization layers with Group normalization layers. Batch normalization layers learn the statistical distribution of signal and background samples during training; however, the distribution of samples during inference doesn't match the training distribution. Changing normalization layers prevents learned batch statistics from impacting performance.
Finally, the last fully-connected network is changed to map to a single sigmoid output, for binary classification.

The training involves using analysis-ready segments of raw detector data
and \emph{timesliding} them to generate a signal-free stream to be used as background or noise. Timesliding involves sliding one
detector greater than the light-travel time between the H1 and L1 sites, and analyzing them coherently~\cite{1989PhRvD..40.3884G}.
This removes any astrophysical correlations, and is an established technique in GW data analysis to generate years-long
astrophysical signal-free stretches from much shorter observing periods (see section III from \cite{s3_and_s4}).
Practically, during training, segments from individual
detectors are sampled independently, constrained to being relatively separated more than the H1-L1 light travel
time, and combined to form coincident inputs.
The use of real data ensures robust learning of non-Gaussian behavior
of the noise and makes the model more resilient to changing background. Notably, we do not use stationary colored
Gaussian noise to generate background as it does not capture the complex non-stationary behavior.
Simulations of CBCs based on general-relativity, referred to as
\emph{injections}, are used on top of the timeslides to create the foreground.
The network is shown $1.5$ seconds of whitened timeseries to perform binary classification i.e., segments containing
the merger of the CBC at any point within the segment are tagged positive, while those that don't are
tagged negative. The whitening procedure uses a power spectral density (PSD) estimated dynamically from preceding data,
which was chosen to be 8-seconds for training and 64-seconds while offline and online inference. This ensures that the model
learns to operate under realistic, non-stationary noise conditions, representative of an LVK observing run. 
Additionally dynamic augmentations, including time reversal, detector swapping, and random muting of signal in
one of the detectors, are done to further expand the negatively tagged training set and enforce the network
to learn coincidence and coherence of signals between the detectors. We note that during training, 
the injected signals are rescaled to a different distribution compared to
an astrophysical population, for example, the training curriculum uses louder injections in the initial
few epochs. However, during validation and testing, the injections are chosen to follow an astrophysical
population to match results reported by the LVK. We also note that much of the preprocessing and signal
projection is implemented within a GPU-accelerated framework~\cite{ml4gw}, eliminating bottlenecks associated with
CPU-based operations and data transfers, and enabling efficient training with greater than 90\% GPU
utilization for our experiments. We provide more details about the dataset preparation, and the
enhancements in Appendix~\ref{appendix:train_val_test}.
The short $1.5$-second kernel using just the
whitened timeseries implies that the network loses sensitivity to signals that are longer than this duration.
Hence, the original {\aframe} sensitivity to BBHs $\gtrsim 25\msun$ is comparable to template-based approaches,
but that for BNSs, where the signals are longer in time, is limited. In the next subsection, we discuss
how we address this limitation.

\subsection{Challenges of Long-Duration Signals for Machine Learning}\label{subsec:ml-limitations}
\begin{savenotes}
\begin{figure}[htp]
    \centering
    \includegraphics[width=1.0\textwidth]{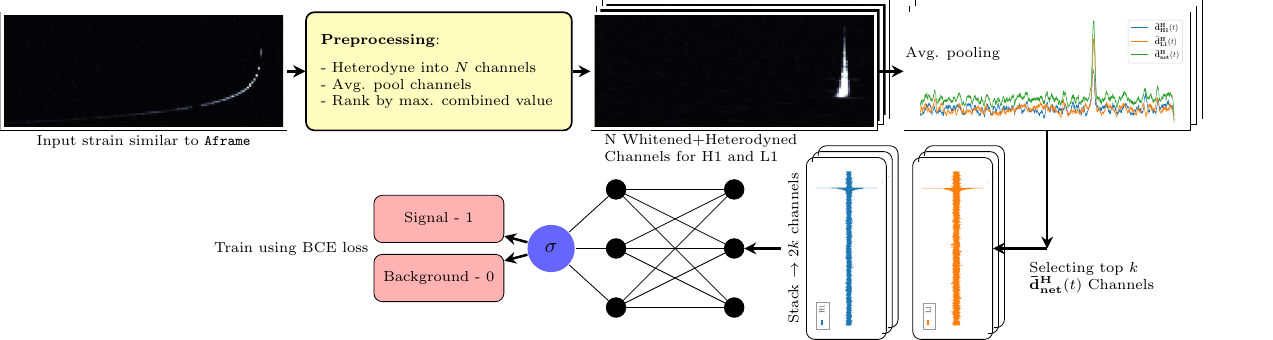} \\
    (a) \\
    \includegraphics[width=0.8\textwidth]{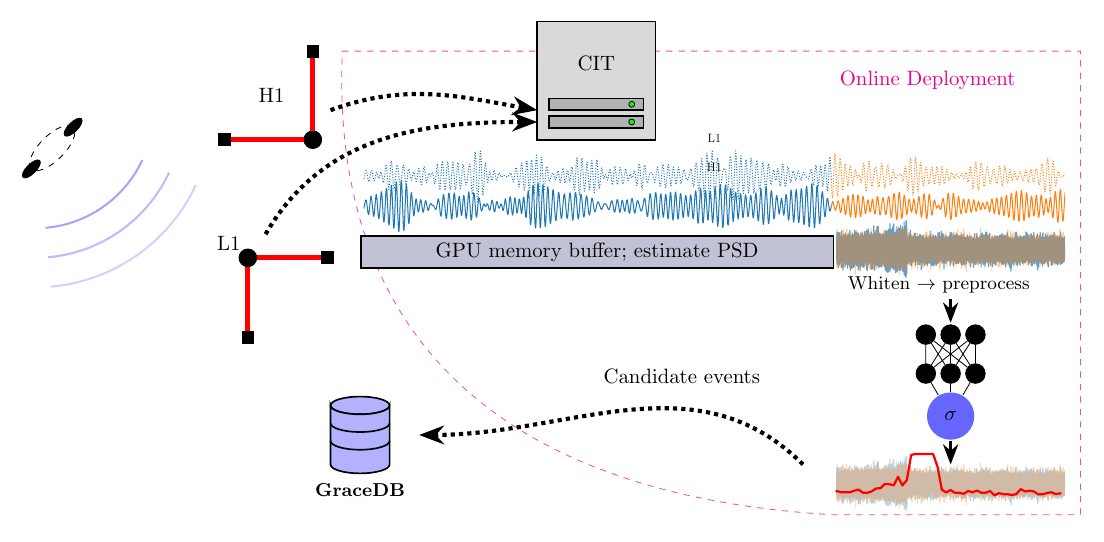} \\
    (b) \\
    \includegraphics[width=0.6\textwidth]{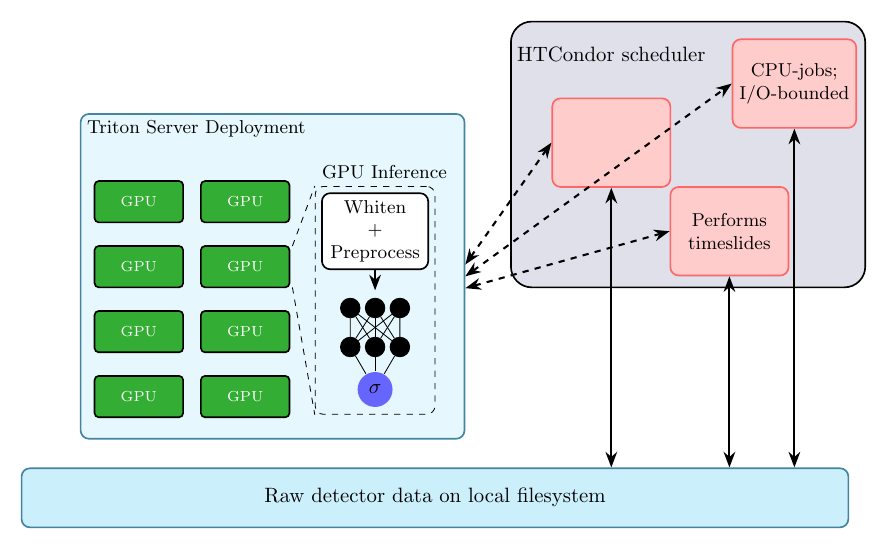} \\
    (c)
    \caption{This figure illustrates the three main components of the {\aframe} search pipeline. (a) \textit{Training pipeline}: detector strain is loaded from disk and processed through a preprocessing stage that includes waveform injection, dynamic power spectral density estimation and whitening, heterodyning, average pooling, and selection of the top $k$ heterodyne channels. The resulting representations are labeled as signal or background and used to train the neural network through binary cross-entropy loss. (b) \textit{Online inference pipeline}: streaming strain data from the H1 and L1 detectors are buffered in a snapshotter and processed using the same preprocessing steps as during training. The selected top $k$ heterodyned channels are passed to the trained neural network, whose outputs are integrated to produce a detection statistic. Events exceeding a public detection threshold are uploaded to GraceDB~\footnote{\url{https://gracedb.ligo.org}} %add citations for gracedb
    for follow-up. (c) \textit{Offline sensitivity analysis}: the trained model is deployed in a streaming configuration using Triton Inference Server. Background distributions are constructed through \emph{timeslides} to estimate false alarm rates, while simulated signal injections are analyzed to measure the sensitive volume. The analysis is distributed across multiple GPUs using {\htcondor} workflow to accelerate large-scale background and foreground processing.
    }
    \label{fig:nn_data_and_model}
\end{figure}
\end{savenotes}

\begin{figure}[h]
    \centering
    \includegraphics[width=0.7\textwidth]{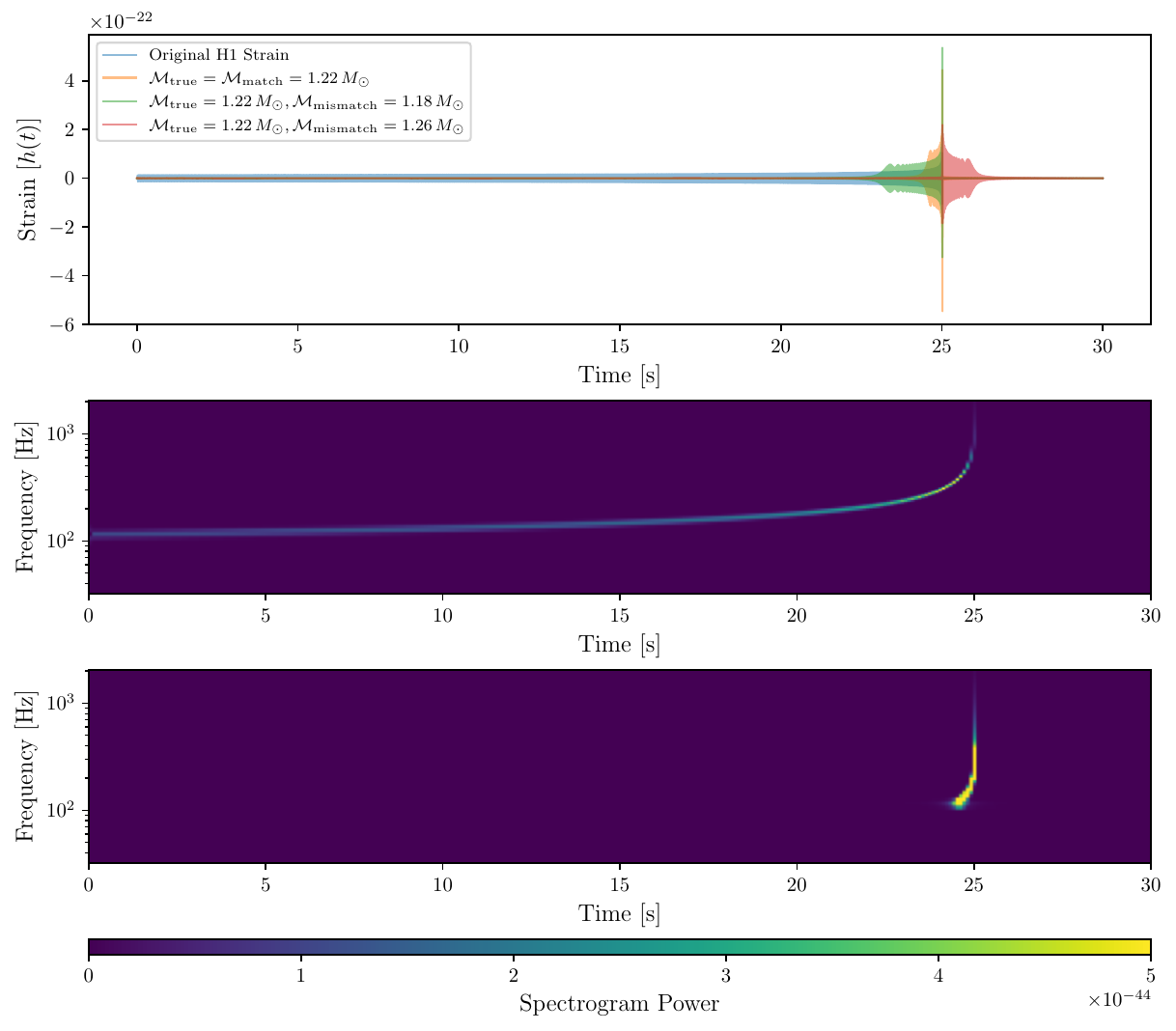} 
    \caption{We show the effect of the heterodyne transformation on a representative binary neutron stars waveform for matched and mismatched chirp masses in the H1 detector. The top panel shows the heterodyned timeseries representation, while the bottom two panels show the corresponding spectrograms. For a matched chirp mass, the inspiral phase is largely removed, and the signal power becomes concentrated near the merger time. This localization is visible in the spectrogram as a concentration of power within a compact temporal region. In contrast, mismatched chirp masses leave a residual phase evolution in the waveform but remain concentrated closer to the merger.}
    \label{fig:timeseries-spectrogram-heterodyne-example}
\end{figure}

Neutron star binaries pose a challenge from a machine-learning standpoint due to the long duration of these signals. For binaries with component masses in the range 1\,--\,2.5\,\msun, signals evolve in the detector frequency band for up to $\mathcal{O}(\mathrm{min.})$. Capturing the full signal would therefore require longer input segments, thereby increasing computational cost and latency.

Existing literature on sequence-based architectures suggests they are often inefficient or difficult to optimize as temporal dependencies grow. For example, in recurrent architectures with long-duration sequences, gradient propagation is reported to be unstable, leading to vanishing or exploding gradients~\cite{bengio1994, pascanu2013}. For convolutional architectures, there may be discrepancies between the theoretical and effective receptive fields. 
Empirically, the effective receptive field occupies only a fraction of the theoretical receptive field, reducing the model's ability to capture temporal dependencies over a range~\cite{luo2017}. Recently, Attention-based Transformers~\cite{vaswani2023} have helped in addressing some limitations, but they still have their computation cost scale quadratically with sequence length, making long inputs expensive and increasing the difficulty of isolating the most informative regions of the sequence. These challenges are relevant for analyzing BNS signals with such architectures. A minute-long signal, sampled at $\mathcal{O}(\text{kHz})$, leads to $\mathcal{O}(10^5)$ tokens! Signal extraction involves picking out \emph{one} signal degree of freedom from $\sim 10^5$ noise degrees of freedom. Naively applying sequence-based architectures to such long-duration signals is both computationally expensive and difficult to optimize effectively given the weakness of the signal and the complexity of the noise.

On the other hand, {\aframe}'s {\resnet} performs effectively when the informative signal content is localized within a short temporal window $\mathcal{O}(\text{seconds})$, like the search for BBHs. This motivates the use of heterodyning as a way to transform long-duration GW signals into shorter, more compact representations in the time domain, where the information is concentrated near the merger.

\subsection{Heterodyne Preprocessing for Long-Duration Signals}\label{subsec:heterodyne}
% \textbf{Heterodyne Preprocessor}: 
GW signals from CBCs are well modeled by general relativity~\cite{MTW1973}.
The frequency-domain waveform is written as,
\begin{equation}
    \tilde{h}(f) = \mathcal{A} f^{-7/6} \exp\left[{i\Psi(f;\boldtheta)}\right],
    \label{eq:spa}
\end{equation}
where $f$ is the frequency, $\mathcal{A}$ is the amplitude factor, $\Psi(f;\boldtheta)$ is the GW phase
from which most of the information about the source is extracted.
The phase depends on the intrinsic parameters of the binary, like masses and spins of the two components,
$\boldtheta = \{m_{1,2}, {\pmb{\chi}}_{1,2}\}$, and is computed perturbatively in the post-Newtonian (PN)
approximation~\cite{blanchet_lrr}. At leading order, or zeroth-PN order, $\Psi$ in eq.~(\ref{eq:spa})
depends on a combination of the component masses, called chirp mass $\mc$ as~\cite{Buonanno_2009},
\begin{equation}
    \Psi_{0\mathrm{PN}}(f;\mc) = \frac{3}{128}\left(\frac{\pi G \mc f}{c^3}\right)^{-5/3}\;\;\text{, with }\mc = \frac{(m_1m_2)^{3/5}}{(m_1 + m_2)^{1/5}}.
    \label{eq:heterodyne_phase}
\end{equation}
This term dominates the evolution of the BNS during its inspiral. If we heterodyne, or demodulate, the signal with a reasonable approximation of chirp mass using eq.~(\ref{eq:heterodyne_phase}), it (a) demodulates the rapidly oscillating signal into a slowly varying one, and (b) concentrates the signal power into a short temporal window where the signal deviates from the inspiral, still preserving the underlying phase information. We, therefore, adopt this approach of de-chirping to transform the signal degree of freedom into a shorter temporal window, where the {\resnet} architecture is efficient in classification. We note that heterodyning has been previously used in GW data analysis literature to remove the rapidly oscillating component of the signal based on a reference phase. Such an approach helps with data compression and fast likelihood evaluations~\cite{Cornish_2021, zackay2018relativebinningfastlikelihood, labrador, dingo}. In this work, we use only phase-based heterodyning and do not apply amplitude weighting, since the primary goal is signal detection. After whitening, phase evolution typically carries most of the information needed to distinguish signals from noise. This can be seen visually in the top panel of Fig.~\ref{fig:nn_data_and_model}, where a spectrogram view of the original and heterodyned versions of a signal are shown. A more detailed view, including the time-domain strain before and after heterodyning is shown in Fig.~\ref{fig:timeseries-spectrogram-heterodyne-example}.

% The heterodyne transform maps the signal into a frame in which the long and rapid oscillatory chirp is demodulated, converting a rapidly evolving waveform into a more slowly varying representation. This effectively converts a chirping signal into a slowly varying representation, allowing long-duration signals to be analyzed within short time windows while preserving the underlying physical information. However, in a search setting, this quantity is also unknown a priori. To address this,

\begin{figure}[h]
    \centering
    \includegraphics[width=0.8\textwidth]{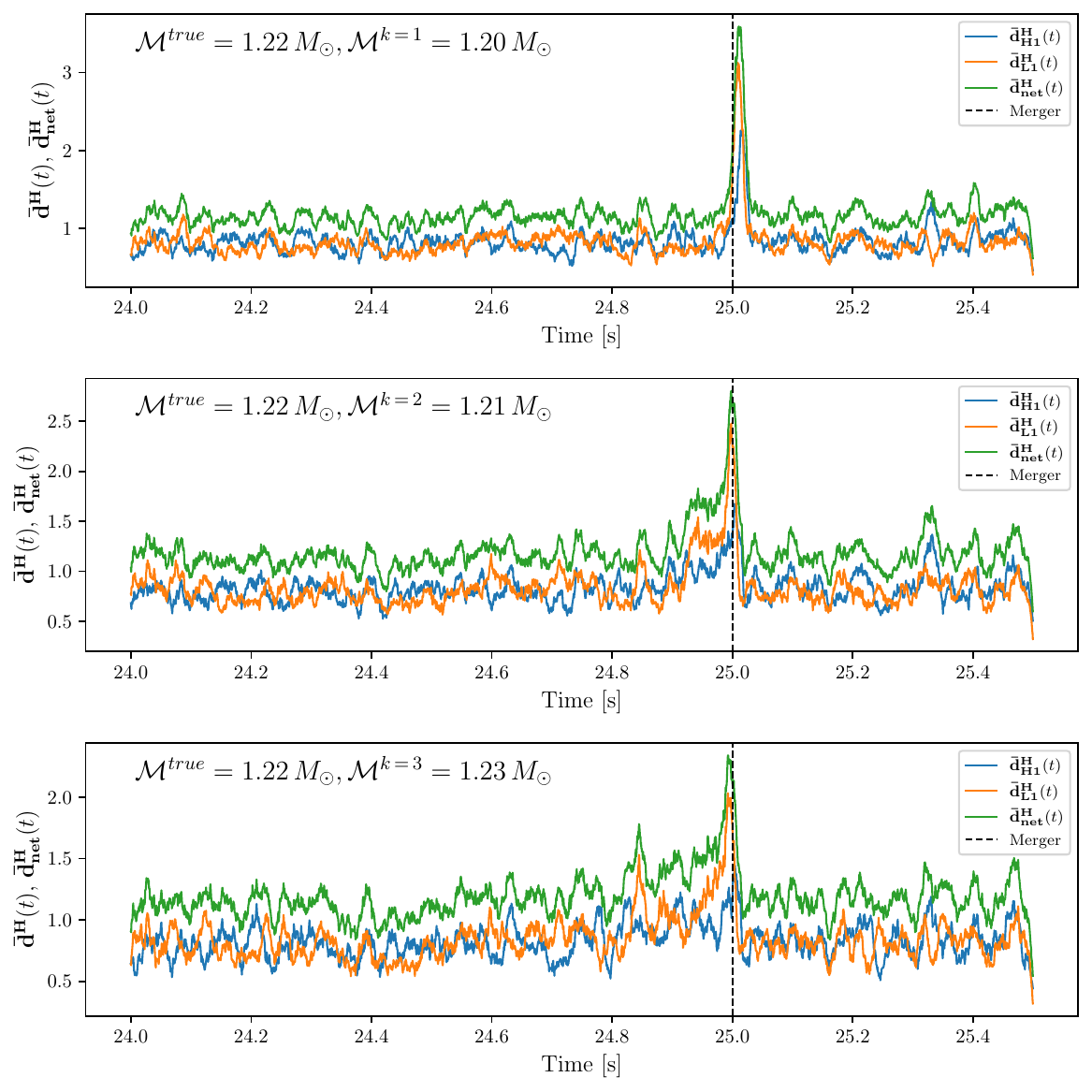} 
    \caption{This figure shows the selection of the most informative heterodyned chirp-mass channels for a binary neutron star signal using a top $k$ ranking based on the peak amplitude of the network-combined response. The panels show the average-pooled heterodyned timeseries {\heterodatapooled}(t) for H1 and L1, for individual chirp-mass channels and the corresponding network-combined representation, $\heterodatapooled_{\mathrm{net}}(t)$. The merger occurs at $t_{c}=25$s, which is near the peak time $t_0 = \argmax [{\heterodatapooled}_{\mathrm{net}}(t)]$ for each of the panels.
    % To enforce coincidence and coherence between detectors, the H1 and L1 pooled responses are combined in quadrature to form {\heterodatapooled}$_{\mathrm{net}}(t)$. 
    Channels are then ranked according to the peak amplitude $\heterodatapooled_{\mathrm{net}}(t)$, and the top $k=10$ channels are retained for subsequent inference. Here the top three are shown for a visual sense.}
    \label{fig:heterodyned-snr-timeseries}
\end{figure}
The challenge for a search algorithm is that the signal (source) parameters are unknown a priori. Therefore, we introduce a bank of $N=100$ chirp mass-dependent reference phases, using eq.~(\ref{eq:heterodyne_phase}), against which we heterodyne the data. For this work, we use a grid spanning 1\,--\,2.5\,\msun, sampled uniformly in $\log \mathcal{M}$. This preprocessing step produces $2N$ heterodyned channels, one per detector for the $N$ reference phases. In the presence of a signal with true (detector frame) chirp mass, $\mc^{\text{true}}$, the channels with $\mc \approx \mc^{\text{true}}$ yield a high-coherence signal in both H1 and L1 channels. This is visually shown in the top panel of Fig.~\ref{fig:nn_data_and_model}. On the other hand, in the absence of a signal, the heterodyning does not introduce any such correlations. In fact, for Gaussian noise, the statistical properties are unchanged if the data is heterodyned by a phase term, like in eq.~(\ref{eq:heterodyne_phase}). 

The heterodyned timeseries, $\heterodata(t)$, using a reference phase $\Psi_{\mathrm{ref}} = \Psi_{0\mathrm{PN}}(f;\mc)$,
is given by
\begin{equation}
    \begin{aligned}
        \heterodata(t) = 2\,\int_{0}^{\infty} \left[\left(\whiteneddata(f)\right)
            \left(e^{-i\Psi_{\mathrm{ref}}(f)}\right)\right]
            e^{2\pi i f t}
            \;df,
    \end{aligned}
    \label{eq:matched_filter}
\end{equation}
where $\whiteneddata(f)$ is the whitened data, collected from the detector, with or without a signal. The integral is the inverse Fourier transform and scans over the different time shifts. In the presence of a signal merging at time $t_c$, the heterodyned timeseries peaks at around $t_c$. The effect is pronounced after taking a running average (average pooling) of the data, as shown in the top panel of Fig.~\ref{fig:nn_data_and_model}.
This technique in eq.~(\ref{eq:matched_filter}) is similar to template-based searches in terms of sliding a template over the data~\cite{findchirp}. We can therefore interpret these $2N$ heterodyned channels as analogs to a sparse template bank over chirp mass. We exploit the coherent peak structure of $\heterodata$ to identify the most informative channels. We enforce coincidence and coherence between detectors by adding the average pooled output in quadrature,
\begin{equation}
    {\heterodatapooled}_{\mathrm{net}}(t) = \sqrt{\vert\heterodatapooled_{\mathrm{H1}}(t)\vert^{2} + \vert\heterodatapooled_{\mathrm{L1}}(t)\vert^{2}}.
    \label{eq:pooling}
\end{equation}

This gives $N$ quadrature channels which we sort based on their peak value, $t_0 = \argmax [{\heterodatapooled}_{\mathrm{net}}(t)]$, from eq.~(\ref{eq:pooling}), and choose the top $k=10$ channels. We take the segment between times $[t_0 - \Delta t, t_0]$ of the heterodyned timeseries data from each detector based on this ranking, i.e., $\{\heterodatapooled_{\mathrm{H1}}, \heterodatapooled_{\mathrm{L1}}\}$ between times $[t_0 - \Delta t, t_0]$, stack them as $2k$ channels, and pass to the neural network for further classification. We use $\Delta t=1.5$ seconds in this work, since most of the coherent power is localized in this region after heterodyning. For comparison, we note that {\aframe} for BBHs used a similar $1.5$-second kernel. Hence, we are able to use an identical {\resnettf} architecture except for having $2k$ input channels. An example of the top three out of $k=10$ channels, in the presence of a signal, is shown in Fig.~\ref{fig:heterodyned-snr-timeseries}. As a summary, the heterodyning method gives an alternate input representation of the data, where long-duration inspiral signals are transformed into shorter ones, and the background is left statistically unchanged. This enables the use of an identical {\resnet} architecture as in the {\aframe} BBH case, which has the discriminative power for signals in this regime. Effectively, this expands the sensitivity of {\aframe} to BNSs just by an intuitive data preprocessing step. This is advantageous in practice, as the benefits of online and offline deployment port over naturally from the BBH case to the BNS case.

\subsection{Training and Validation Procedure}\label{subsec:training}
\begin{table}
    \centering
    \begin{tabular}{llll}
        \hline
        Parameters                      & Priors                              & Limits           & Units  \\
        \hline
        Mass of primary                 & $m_{1} \sim$ Triangular(mode = 2.5) & (1, 2.5)         & \msun  \\
        Mass of secondary               & $m_{2}$                             & (1, $m_{1}$)     & \msun  \\
        Redshift                        & Comoving                            & (0, 0.15)        & -      \\
        Polarization angle              & Uniform                             & (0, $\pi$)       & rad.   \\
        Dimensionless spin magnitude    & Uniform                             & (0, 0.4)         & -      \\
        Spin tilt                       & Sine                                & (0, $\pi$)       & rad.   \\
        Relative spin azimuthal angle   & Uniform                             & (0, $2\pi$)      & rad.   \\
        Spin phase angle                & Uniform                             & (0, $2\pi$)      & rad.   \\
        Orbital phase                   & Uniform                             & (0, $2\pi$)      & rad.   \\
        Right ascension                 & Uniform                             & (0, $2\pi$)      & rad.   \\
        Declination                     & Cosine                              & ($-\pi/2$, $\pi/2$) & rad.   \\
        Inclination angle               & Sine                                & (0, $\pi$)       & rad.   \\
        \hline
    \end{tabular}
    \caption{Distribution of the different priors used to generate the training, validation, and testing waveforms. These priors are taken from GWTC-3~\cite{Abbott_2023} to compare with the search pipelines.}
    \label{tab:simulation params}
\end{table}

\subsubsection{Data}\label{subsec:training-data}

We use publicly available strain data from LVK's O3 observing run from the Gravitational Wave Open Science Center (GWOSC)~\footnote{\url{https://gwosc.org/}}~\cite{gwosc3}. The data is resampled to 2048\,Hz, and segments that are coincident and analysis-ready between H1 and L1 are retained. In particular, training uses data between 2019-11-01 to 2020-01-05, with 15,000\,seconds reserved for validation. For evaluating the performance of the model, we use a separate segment, non-overlapping with training background, between 2019-05-09 to 2019-06-08, and apply the same coincidence and data-quality criteria. The use of non-overlapping stretches mimics a real-time scenario in which the neural network must generalize to different segments of detector strains without prior exposure to a certain period.

\subsubsection{Waveforms}\label{subsec:waveforms}

We generate 30,000 BNS waveform polarizations, with 25\,s of duration sampled at 2048\,Hz using {\bilby} with the {\phenomp} approximant, of which 10,000 are used for training, and 20,000 are used for validation. The intrinsic and extrinsic source parameters for the waveforms are drawn from astrophysical distributions consistent with those used in GWTC-3 sensitivity studies, mentioned in Table\,\ref{tab:simulation params}. 
During training, the extrinsic projection parameters are sampled on-the-fly, enabling real-time data augmentations and exposing the network to a variety of signal realizations. For testing, waveforms are generated using the same distribution (See Table\,\ref{tab:simulation params}). The source population is distributed uniformly in co-moving volume where the majority of signals are intrinsically low signal-to-noise ratio (SNR).
Instead of injecting signals that are not expected to be recovered at any reasonable detection threshold, we adopt an SNR-based rejection sampling strategy, retaining only those injections with network $\mathrm{SNR}>4$, spacing them out adequately to avoid overlap. We, however, account for all generated samples in estimation of sensitivity (See Section III in \cite{aframe-methods} and Section \ref{sec:performance} below). This reduces variance in the sensitivity estimate without biasing the result, and is a standard practice followed in reporting search sensitivities by the LVK~\cite{Abbott_2023}. For this work, we sampled $\sim 4\times10^5$ signals, but generated and injected $\sim 8\times10^4$, the rest being rejected by the $\mathrm{SNR}$ threshold for ``hopeless'' injections.

\subsubsection{Training Strategy}\label{subsec:training-strategy} 

A training batch is constructed dynamically by loading 30-second segments of raw detector data from H1 and L1 with timeslides (see section~\ref{subsec:heterodyne}), giving 2-channels input. For every element, the first 20\,s are used to calculate a PSD per-channel, that is used to whiten the remaining segment. Simulated waveforms are injected in the remaining segment for a random fraction of the batch. Following this the remaining 10\,s is whitened using the estimated PSD. To account for the whitening filter's edge effects, 1\,s of data is trimmed from both edges leaving a batch of 8\,s of whitened data, \whiteneddata. As mentioned above, for injections, the waveform polarizations are projected onto the detector responses on-the-fly and rescaled to the target SNR, thereby increasing the diversity of signal realizations encountered during training. Also, the coalescence time is randomized within 1\,s from the right edge of the analysis window to enforce approximate time-translation invariance and expose the network to signals arriving at different locations of the segment. Additional data augmentations include time-reversal and amplitude inversion for the raw data, and swapping the signal between H1 and L1 or muting the injection in one of the channels. These are done with fixed probabilities to improve robustness to varied noise morphologies and transient artifacts, and to learn the concept of signal coincidence and coherence. The batch elements are tagged positively or negatively depending on the elements containing injections. Those entries that involve swapping or partially muting an injection are tagged negatively. These augmentations are identical to the training strategy used for the BBH variant of {\aframe}.

Following injection and whitening, the whole batch containing both positive and negative elements is heterodyned as described in section~\ref{subsec:heterodyne}. Operationally, this involves mapping the batch of whitened time-domain to the frequency domain via an FFT operation. Here, the data is heterodyned using $N$ reference waveforms into $2N$-channels, and converted back into time-domain via iFFT, giving $\heterodata$ using eq.~(\ref{eq:matched_filter}). The channels are average-pooled, giving $\heterodatapooled$. This is then sorted in descending order using eq.~(\ref{eq:pooling}) and restricted to the top $2k$ channels. Lastly, the rightmost $\Delta t=1.5$ seconds for each element is sliced since it is sufficient for classification. Thus, the preprocessed batch that is input to the neural network involves 1.5-second kernels with $2k$-channels of heterodyned timeseries. This network is trained for binary classification using the binary cross-entropy loss.

\subsubsection{Validation Strategy}\label{subsec:validation-strategy}  

We adopt the same validation prescription as used in {\aframe} BBH search \cite{aframe-methods}. Segments of detector strains following the training period, along with 20,000 simulated waveforms, are reserved for validation. The validation set is prepared similarly as described in \cite{aframe-methods}. This involves adding simulated signals into noise segments, network-SNR thresholding, rescaling, and multiple coalescence time placements, ensuring close agreement between validation and testing conditions. Pre-processing of the validation set mimics the training setup; each segment is dynamically whitened using PSD estimates from preceding data, followed by the heterodyning of the whitened strain. The final 1.5-seconds of the heterodyned strains are forwarded to the network for analysis. Validation performance is measured using the area under the receiver operating characteristic curve (AUROC) evaluated up to a false positive rate of $1\times10^{-3}$, emphasizing the low false-alarm regime relevant for gravitational-wave searches. The model checkpoint achieving the highest validation score is selected for final testing.

\section{Comparison of Search Sensitivity Across Existing Pipelines}\label{sec:performance}
\subsubsection{Sensitive Volume}\label{subsec:sensitive-volume} 

One of the key metrics for evaluating the performance of GW search
pipelines is its astrophysical sensitive volume at a fixed false alarm rate (FAR). Typical
threshold values for considering a candidate of interest is FAR $\sim 1/\text{month} \approx 3.9\times10^{-7}\;\text{Hz}$,
which was used for sending public alerts in LVK O4~\cite{emfollowupguide}.\footnote{Another representative FAR threshold
$=1/\text{year}\approx 3.2\times10^{-8}$ is used for offline LVK data analysis~\cite{gwtc4}.}
A candidate with this FAR value indicates the pipeline can find no more than
one such \emph{background} candidate in one-month-long data analysis. Hence, a smaller FAR value implies greater significance.
Therefore, sensitive volume, as a function of FAR, provides an astrophysically meaningful measure of search sensitivity by 
quantifying the effective volume of the Universe within which a pipeline can detect binary mergers. 
It incorporates both the pipeline's detection efficiency and the underlying astrophysical distribution of sources,
providing a direct connection to detection rates. In practice, it is estimated using Monte Carlo
integration by drawing signals from an assumed, most accurate astrophysical population, injecting them into real detector
noise, and measuring the fraction recovered by a search algorithm at a given FAR threshold. If the injections are drawn
uniformly in co-moving volume, the sensitive volume can be approximated as the total surveyed volume multiplied
by the fraction of recovered signals. To evaluate sensitivity to certain representative systems, importance
sampling is employed to re-weight the injections from a broad distribution,
% allowing the sensitive volume to be computed for a subset of the population 
weighting them by the ratio of the target to the sampling distribution. To reduce the statistical uncertainty
in the sensitive volume calculations, we use SNR-based rejection sampling, where waveforms with $\mathrm{SNR}<4$
are excluded as they are not expected to be recovered at any relevant FAR and therefore, contribute negligibly.
However, these samples are still counted in the total number of draws, effectively increasing the sampling
efficiency without biasing the estimate. Here we re-weight the same astrophysical population as used in GWTC-3
sensitivity studies, where a log-normal distribution around representative binary masses, $\hat{m}_{1,2}$,
with a width $\Delta m_{1,2}=0.1\msun$ was used. Furthermore, detections are required to occur within 1\,s of
the true coalescence time, consistent with the time resolution of downstream analyses, like parameter estimation\cite{aframe-methods}.

\subsubsection{Background Estimation}\label{subsec:background-estimation}  

The FAR quantifies the expected rate at which random noise fluctuations produce triggers with a detection statistic exceeding a given threshold. These values are estimated empirically from the background distribution obtained using timeslides
that generate a signal-free stream of several years (see section~\ref{subsec:bbh}). While
constructing timeslides, we keep the H1 detector strain fixed while the L1 strain is incrementally shifted to accumulate the required background livetime. For our results, we accumulate 10 years of livetime. Using the constructed background dataset, the FAR of a candidate event is computed by counting how often parts of this signal-free stream produces events with a detection statistic at least as large as that of the
candidate, and normalizing by the total background livetime. We emphasize that no detector strain used for this FAR calculation is included in the training or validation sets. 
We further penalize our foreground candidate significance since the search presented here only spans the BNS mass range, as opposed to
the full parameter space of NSs and BHs. Since we operate a BBH-only search in addition, it can be considered as another \emph{trial}
in accumulating background candidates. Therefore, we multiply our FAR values from foreground using a trials factor of 2 to count two different searches
operating together. Thus, although 10 years of background livetime are computed, the lowest FAR reported in this work is
2 per 10 years, or equivalently, $1/5$-years. We note that this is a conservative penalty since we observe substantial overlap between
our BBH and BNS background populations, indicating that the two searches are not statistically independent, i.e., in practice the trials
factor $< 2$. However, to facilitate a more meaningful comparison with matched-filter pipelines, which operate in the mass range including
both NS and BH compact-objects, we conservatively apply this penalty to our BNS FAR estimates.

\subsection{Comparison with GWTC-3 Matched-Filter Searches}\label{subsec:mf-comparison}
\begin{figure}[h]
    \centering
    \includegraphics[width=1.0\textwidth]{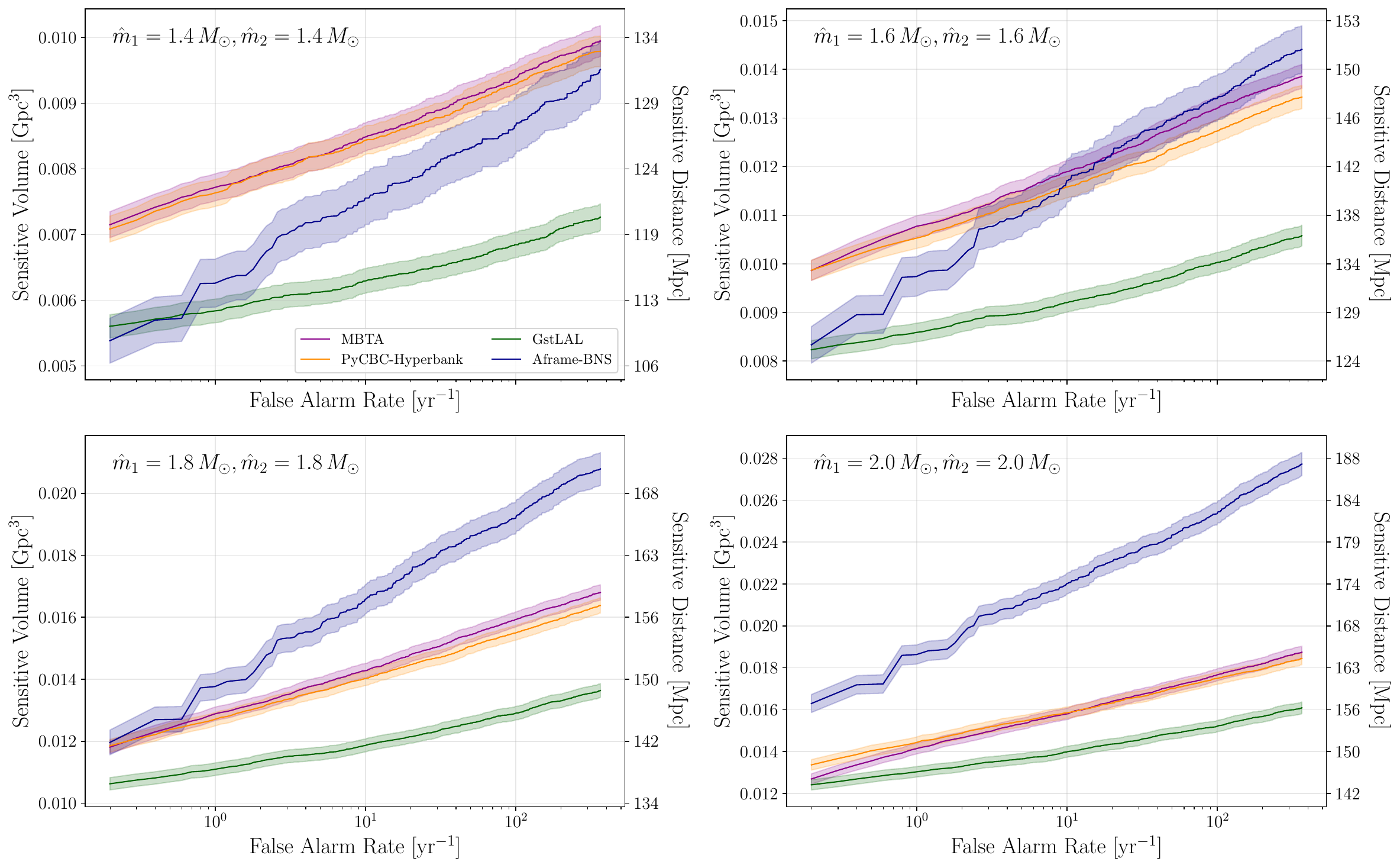}
    \caption{Sensitive volume as a function of false alarm rate for four representative binary neutron star mass distributions:
    $\hat{m}_{1,2} = \{1.4, 1.6, 1.8, 2.0\}\;\msun$ in the source frame, with a width of $\Delta m_{1,2}=0.1\msun$.
    For each representative system, component masses are drawn from log-normal distributions centered on the quoted masses.
    The re-weighting procedure and the sensitivities for matched-filter pipelines are taken from GWTC-3~\cite{Abbott_2023} estimates.
    We show that {\aframe} achieves sensitivity comparable to matched-filter searches.}
    \label{fig:sensitive-volume}
\end{figure}

We benchmark {\aframe}'s performance using the sensitive volume for certain representative BNS masses, enabling a direct astrophysical comparison with low-latency matched-filter search pipelines used within the LVK collaborations. Fig.\,\ref{fig:sensitive-volume} compares the sensitive volume of {\aframe} as a function of FAR with those reported for matched-filter searches: \mbta, \gstlal, and \pycbchyperbank, in GWTC-3~\cite{Abbott_2023}. The FAR values shown for {\aframe} include the trials-factor correction described above, corresponding to a factor-of-two penalty applied to facilitate direct comparison with all-sky compact-binary searches. We see that {\aframe}'s sensitive volume is comparable to that from match filtering pipelines across the BNS parameter space.
% For lower-mass binaries, such as 1.4\,{\msun}\,--\,1.4\,{\msun} and 1.6\,{\msun}\,--\,1.6\,{\msun} representative populations, {\aframe} achieves sensitivity comparable to matched-filter pipelines. 
The differences in sensitivity that are observed are similar to the variations seen among the matched-filter pipelines. A slight reduction in {\aframe}'s sensitivity is observed at $\mathrm{FARs}<\mathrm{1/year}$ for lower-mass binaries, such as 1.4\,{\msun}\,--\,1.4\,{\msun}, where the matched-filter searches retain an advantage. However, for higher-mass BNS systems, including the 1.8\,{\msun}\,--\,1.8\,{\msun} and 2.0\,{\msun}\,--\,2.0\,{\msun} binaries, {\aframe} exceeds the sensitive volume achieved by the matched-filter pipelines. This trend suggests that the heterodyned representation is particularly effective at concentrating the inspiral-merger structure of shorter-duration, higher-mass binaries into compact temporal features that are easier for the network to
identify. Since these systems are of shorter duration, a larger fraction of their recoverable SNR is contained within the 8-second temporal windows processed by the network.

The decrease in sensitivity toward the lighter end of the BNS mass values can partly %{primarily} 
be attributed to the current design choice of applying heterodyning to only 8-second kernels. Low-mass BNS systems remain in the detector band for much longer durations, causing a significant fraction of their inspiral SNR to fall outside the analyzed segment. By comparison, higher-mass systems evolve more rapidly and accumulate a larger fraction of their recoverable SNR within the 8-second interval, naturally favoring the current setup. Extending the heterodyning procedure to longer input durations could therefore further improve sensitivity to low-mass systems while retaining the advantages of the compressed heterodyned representation. Despite that, achieving performance comparable to matched-filter pipelines in the BNS parameter space demonstrates that the heterodyned pre-processing captures a substantial fraction of the inspiral information within a compact representation suitable for neural-network inference.

\subsection{Comparison with Existing Machine Learning-Based Searches}\label{subsec:pycbc_ml_paper}
While many previous machine-learning searches for GWs report performance primarily on receiver operating characteristic (ROC) curves, this metric is not directly astrophysically interpretable because it depends on the assumed distributions of simulated signals rather than the astrophysical distribution and noise modeling used to construct the evaluation set. One exception is the machine learning-based BNS search algorithm described in \citet{schafer2020}, which evaluates performance using sensitive distance and reports a sensitivity of 50\,{\mpc} at a FAR of $\sim 1/\text{month}$ for NS binaries with component masses between 1.2\,--\,1.6\,{\msun} (See Fig.\,4 in \cite{schafer2020}). We therefore use this result as a point of comparison for our {\aframe} BNS search.

\begin{savenotes}
\begin{figure}[h]
    \begin{subfigure}{0.4\textwidth}
        \centering
        \includegraphics[width=\linewidth]{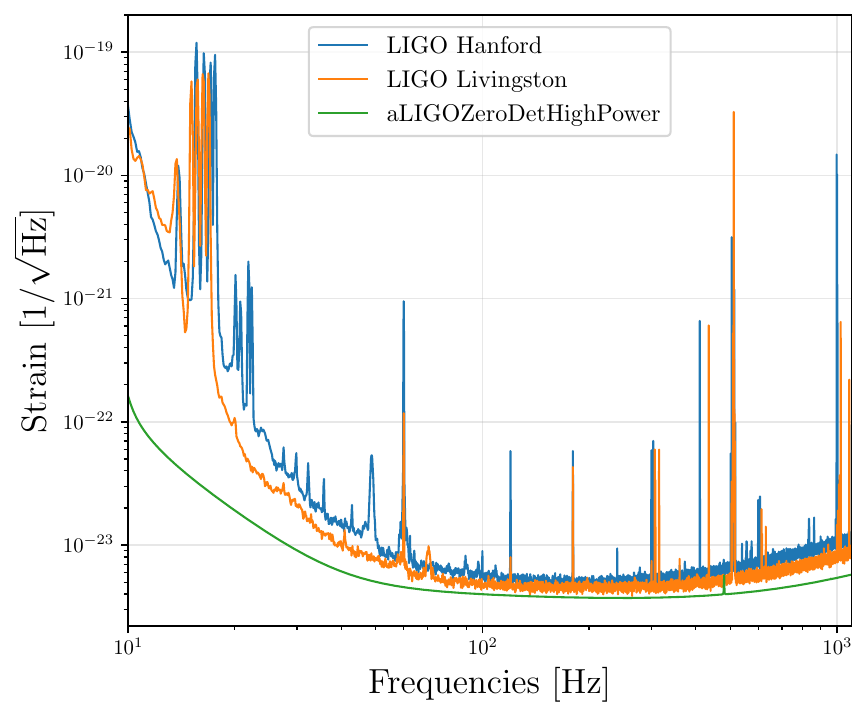}
        \caption{}
        \label{fig:detector-idealized-strain}
    \end{subfigure}
    \begin{subfigure}{0.48\textwidth}
        \centering
        \includegraphics[width=\linewidth]{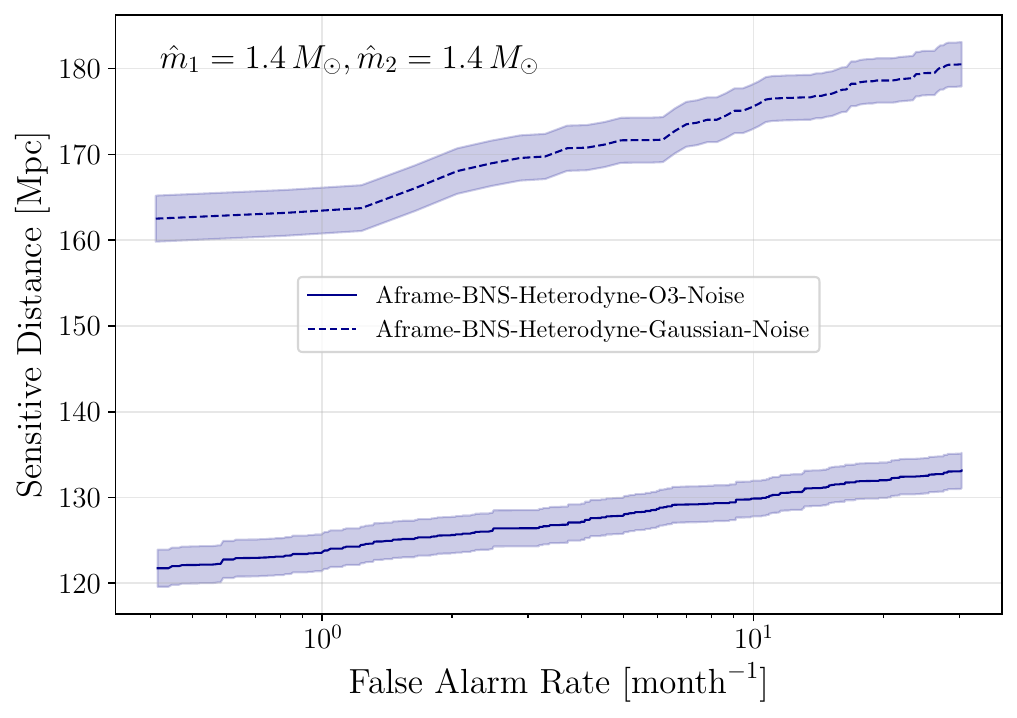}
        \caption{}
        \label{fig:sensitive-distance}
    \end{subfigure}
    \caption{
    (a) Amplitude spectral density (ASD) of the LIGO detectors during the O3 observing run,
    compared to the idealized Gaussian noise {\aligonoise}~\footnote{\url{https://dcc.ligo.org/LIGO-T0900288-v3/public}}$^{,}$\footnote{\url{https://dcc.ligo.org/LIGO-T070247-v1/public}}%, \url{https://dcc.ligo.org/LIGO-T070247/public}
    , in \lalsim~\cite{lalsuite}. The sky-averaged binary neutron star inspiral range for 1.4\,--\,1.4\,{\msun} systems is estimated to be 110\,--\,140\,{\mpc} for the LIGO Hanford~\footnote{\url{https://dcc.ligo.org/LIGO-G1401390/public}}\,
    and LIGO Livingston~\footnote{\url{https://dcc.ligo.org/LIGO-G1500622/public}}\, detectors and 180\,{\mpc} for the {\aligonoise} strain.
    (b) Sensitive distance as a function of false alarm rate for 1.4\,--\,1.4\,{\msun} (source frame)
    binary neutron star systems. Component masses are drawn from log-normal distributions centered
    on the quoted masses, $\hat{m}_{1,2} = 1.4{\msun}$, with a width of $\Delta m_{1,2}=0.1\msun$.
    Solid curves denote sensitivities evaluated using real detector noise from the O3 observing run,
    while dashed curves correspond to evaluations using idealized Gaussian noise, {\aligonoise}.
    At a FAR$\sim 1/\text{month}$, the sensitive distance in case of {\aligonoise} is $\sim 160\mpc$,
    which is three times the value reported at the same FAR, and using the same noise curve in \citet{schafer2020}.
    The sensitive distance for O3 at the same FAR is $\sim 120\mpc$ which is consistent with the
    inspiral range reported for O3.
    }
\end{figure}
\end{savenotes}

To enable a direct comparison between the two approaches, we evaluate {\aframe} under comparable conditions. The \citet{schafer2020} model is trained and evaluated on simulated Gaussian noise corresponding to the {\aligonoise} noise model in {\lalsim}~\cite{lalsuite}, which represents an idealized detector sensitivity free of non-stationary noise transients and instrumental artifacts, as shown in Fig.\,\ref{fig:detector-idealized-strain}. The BNS range, for canonical 1.4\,--\,1.4\,{\msun} NS binaries producing network $\mathrm{SNR}>8$ averaged over the sky, associated with {\aligonoise} is $\sim180\,{\mpc}$.
We perform inference using our model trained on O3 noise, on injections done this {\aligonoise} noise model. We find that at a FAR of $\sim 1/\text{month}$, {\aframe} achieves a sensitive distance of 160\,{\mpc}, indicating three times improved sensitivity relative to \citet{schafer2020}, which reports $\sim 50\;\mpc$ at a FAR of $\sim 1 / \text{month}$. The variation of sensitive volume with FAR is shown in
Fig.\,\ref{fig:sensitive-distance}. Notably, our result is achieved despite {\aframe} being trained on real detector noise from the O3 observing run, highlighting the model's robustness and ability to generalize across different noise conditions. The higher sensitivity obtained using {\aligonoise} is primarily because its amplitude spectral density (ASD) corresponds to a significantly more sensitive detector than the O3 instruments (See Fig.~\ref{fig:detector-idealized-strain}.) As a result, signals produce larger network SNRs at a given distance, allowing them to be detected farther out in space.
Also, the stationary Gaussian noise contains fewer non-Gaussian transients than the real detector data, resulting in a cleaner background distribution. 

For reference against real O3 detector noise, we convert our result of sensitive volume in the previous section into sensitive distance,
defined as the radius of the sphere that encloses the same volume, and overlay it in Fig.\,\ref{fig:sensitive-distance}.
We find that in this realistic setting, {\aframe} achieves a sensitive distance of $\sim$120\,\mpc\, for 1.4\,--\,1.4\,{\msun} system
at a FAR of $\sim 1/\text{month}$. This is consistent with the BNS range, defined for canonical 1.4\,--\,1.4\,{\msun} binaries,
reported in O3 as 110\,--\,140\,{\mpc} (see Fig.\,3 in \citet{Abbott_2023}), which further validates our performance.

In addition to differences in noise modeling, the two approaches also differ substantially in the preprocessing and input representations. The model described in \citet{schafer2020} operates on longer input segments ($\sim 32$\,s), which are further divided into smaller, frequency-dependent segments based on multi-banding and processed by separate neural networks in parallel, with the resulting outputs combined to obtain a final detection statistic. This allows the model to learn a larger fraction of the inspiral evolution directly. In contrast, {\aframe} operates on shorter input windows ($\sim 8$\,s) augmented through chirp-mass-dependent heterodyning. The network receives a compact multi-channel representation in which the long inspiral structure has been localized near the merger, and only the final 1.5\,s of the heterodyned strain are provided to the network. It reduces the burden of learning long-range temporal dependencies while preserving sensitivity to low-mass systems. Finally, the comparison between the two approaches demonstrates that {\aframe} achieves improved sensitivity relative to prior machine learning-based BNS searches under both realistic and Gaussian-noise conditions, highlighting robustness and strengthening the case for deploying machine learning-based approaches alongside traditional pipelines in future observing runs.

\subsection{Long-Term Stability of the Search Pipeline}\label{subsec:longevity}
Any search algorithm intended for real-time GW candidate detection must maintain consistent performance as detector conditions evolve throughout an observing run.
One of the primary challenges is the non-stationary nature of the background caused by a variety of
reasons: environmental conditions, calibration updates, transient instrumental artifacts, and other technical sources of noise~\cite{DetChar-O3}. 
Consequently, it is important to assess the stability of a trained model under these evolving conditions and determine whether periodic retraining is required.
\begin{figure}[h]
    \centering
    \includegraphics[width=0.9\textwidth]{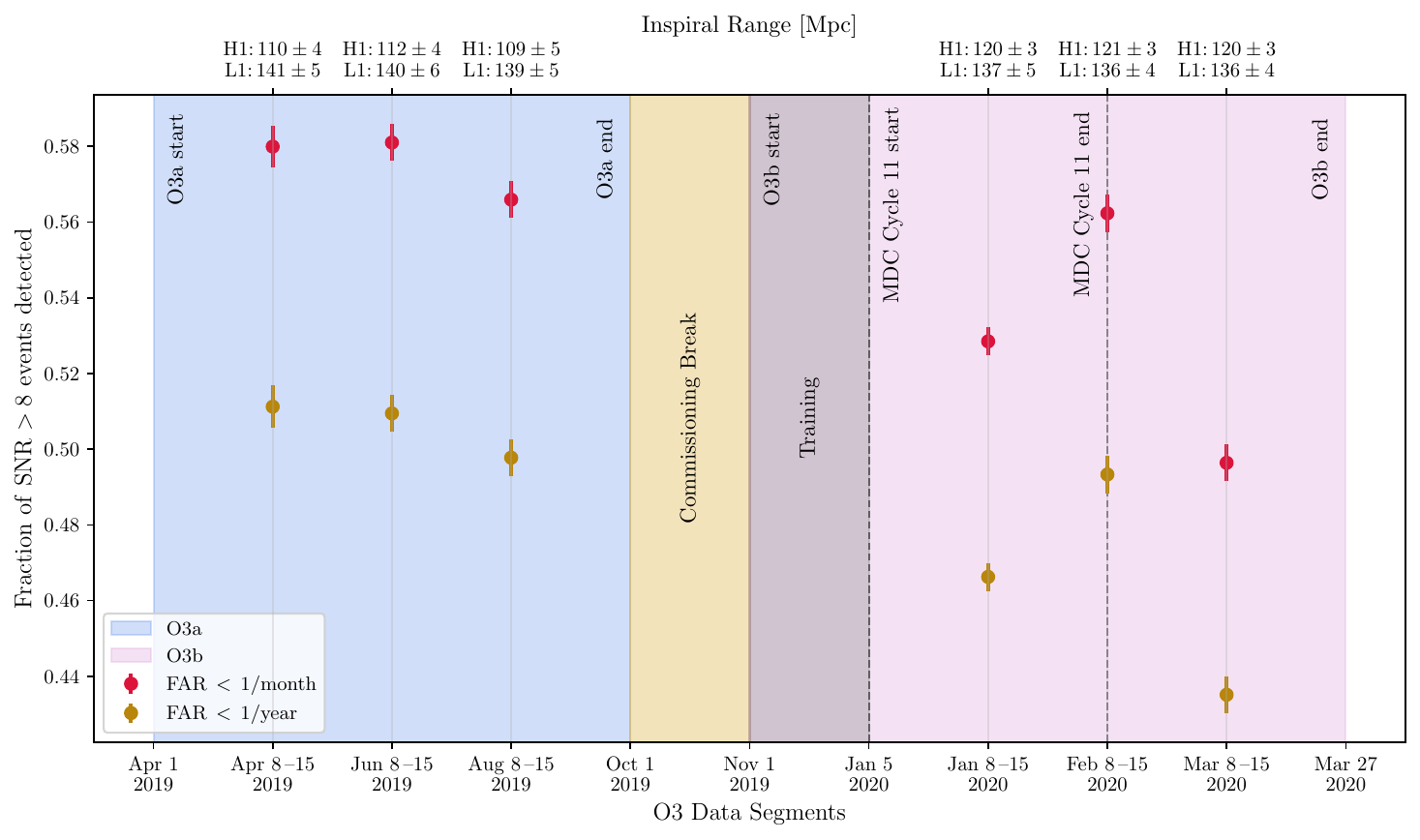} 
    \caption{The fraction of network $\mathrm{SNR}>8$ injected events recovered at false alarm rate thresholds of $\mathrm{1/month}$ and $\mathrm{1/year}$ for different weeks throughout the O3 observing run.}
    \label{fig:longevity}
\end{figure}

For our main result in Fig.~\ref{fig:sensitive-volume}, we performed inference on data from the first half of the O3 run (referred to as O3a) for a model trained using background from the second half of the same run (O3b). We take this a step further and perform injections in other several segments across O3, and
measure the fraction of injected signals with network $\mathrm{SNR}>8$ that are recovered at or below FAR thresholds of $1/\mathrm{month}$ and $1/\mathrm{year}$.
The FAR thresholds used in this analysis are derived from the same background timeslides employed in the sensitive volume calculation shown in Fig.~\ref{fig:sensitive-volume} i.e. from O3a, while the model is trained on O3b. This isolates the test to the case where neither the model, nor the background evaluation have overlap with the testing periods i.e., running live in an observing run.
The results, shown in Fig.~\ref{fig:longevity}, indicate that the recovery fraction remains broadly stable throughout the observing run. While performance exhibits week-to-week fluctuations, the overall trends are consistent across both FAR thresholds. We observe slightly higher recovery fractions during the early stages of O3a and a modest decrease toward the end of O3b, with the lowest performance occurring near the conclusion of the run. The magnitude of these variations is small compared to the overall sensitivity of the search, suggesting that the network generalizes well across a broad range of detector conditions.
In addition, this trend is broadly consistent with the evolution of the detector sensitivities over the observing run. We observe the inspiral range of the more sensitive interferometer (L1) decreases across the selected weeks and reaches its minimum value during the final week, coinciding with the lowest recovery fraction. Although the less sensitive interferometer (H1) exhibits a modest increase in inspiral range over the same period, the overall network performance appears to be primarily driven by the detector contributing the largest fraction of the network SNR. The observed variation is therefore consistent with changes in detector sensitivity rather than a degradation in the network's ability to generalize to evolving noise conditions. As a result, this indicates that a model trained on a limited portion of an observing run can remain effective over timescales of $\lesssim 1$ year, implying that retraining is unlikely to be required on monthly timescales for the detector conditions considered here. Even in the case that retraining is involved, the model can be finetuned from an existing checkpoint on new data collected, as opposed to training from scratch, and hence significantly faster.

% \section{Parameter Estimation using AMPLFI}\label{sec:amplfi}
% \input{sections/amplfi}

\section{Analysis on the O3 Mock Data Challenge}\label{sec:mdc_results}
In order to ensure further consistency in our results, and compare against established match-filter searches used within in LVK real-time analysis, we run our algorithm on the O3 mock data challenge (MDC) dataset~\cite{Chaudhary:2023vec}. This MDC was generated as a part of testing the low-latency alert infrastructure of the LVK before the fourth observing run (O4). It involves CBC injections, including BNSs, done over a stretch of O3b data separate from both our training and testing stretches (See Fig.~\ref{fig:longevity} for times). However, the maximum mass of NSs in this dataset was $2.05\msun$ corresponding to the maximum mass supported by the SLy equation of state~\cite{sly}. Also, the waveforms injected for them included tidal effects in NSs with the {\phenompnrt} waveform approximant. Our waveform approximant during training does not include tidal effects. Hence, this dataset serves as a good benchmark both in terms of changing background and also waveform systematics, though the latter is expected to be small in this context.
Table\,\ref{tab:o3_mdc_events} summarizes the number of recovered injections in the O3 MDC dataset for different search pipelines at FAR thresholds of $\mathrm{1/month}$ and $\mathrm{1/year}$ Across the injections with mass range $1 < m_{1,2} < 2.05$\,{\msun} and network $\mathrm{SNR} > 8$, {\aframe} recovers comparable injections, and more in some cases, than low-latency matched-filter pipelines {\pycbc} and {\spiir} and achieves performance comparable to {\mbta} and {\gstlal}. 

\begin{table}[h!]
\centering
\begin{tabular}{l@{\hspace{0.5cm}}c@{\hspace{0.5cm}}c@{\hspace{0.5cm}}c@{\hspace{0.5cm}}c@{\hspace{0.5cm}}c@{\hspace{0.5cm}}c}
\toprule

& \multicolumn{3}{c}{FAR $<{\mathrm{1/month}}\,(3.8{\times}10^{-7}\,\mathrm{Hz})$}
& \multicolumn{3}{c}{FAR $<{\mathrm{1/year}}\,(3.2{\times}10^{-8}\,\mathrm{Hz})$} \\

\cmidrule(lr){2-4}
\cmidrule(lr){5-7}

Pipeline
& $2.05 > m_{1,2} >1.0$
& $m_{1,2} > 1.7$
& $m_{1,2} > 1.9$
& $2.05 > m_{1,2} >1.0$
& $m_{1,2} > 1.7$
& $m_{1,2} > 1.9$ \\

\midrule

{\mbta}   & 648 & 131 & 19 & 619 & 124 & 17 \\
{\gstlal} & 528 & 105 & 15 & 497 & 99 & 14 \\
{\spiir}  & 462 & 96 & 13 & 414 & 83 & 10 \\
{\pycbc}  & 286 & 59  & 6 & 187 & 36  & 3  \\
{\aframe} & 671 & 142 & 34 & 572 & 123 & 30 \\

\bottomrule
\end{tabular}
\caption{O3 MDC detected event counts for different search pipelines at FAR thresholds of $\mathrm{1/month}$ and $\mathrm{1/year}$. There are a total of 1790 injections chosen with the criteria: $1 < m_{1,2} < 2.05$\,{\msun} and network $\mathrm{SNR} > 8$. 
% Note that these counts are reported without applying a trials-factor correction. 
Note that the FAR values for {\aframe} include a penalty of a factor of 2 assumed due to a trials factor correction.
The FAR values for the other search pipelines are directly taken from the individual event candidates reported in a testing instance
of GraceDB that was used for the O3 MDC benchmarking exercise reported in \citet{Chaudhary:2023vec}.
% Accounting for the trials factor across CBC searches would modify the effective false alarm rate thresholds and consequently the number of events exceeding those thresholds.
}
\label{tab:o3_mdc_events}
\end{table}

\begin{figure*}[h!]
    \centering
    % Top row
    \begin{subfigure}{0.48\textwidth}
        \centering
        \includegraphics[width=\linewidth]{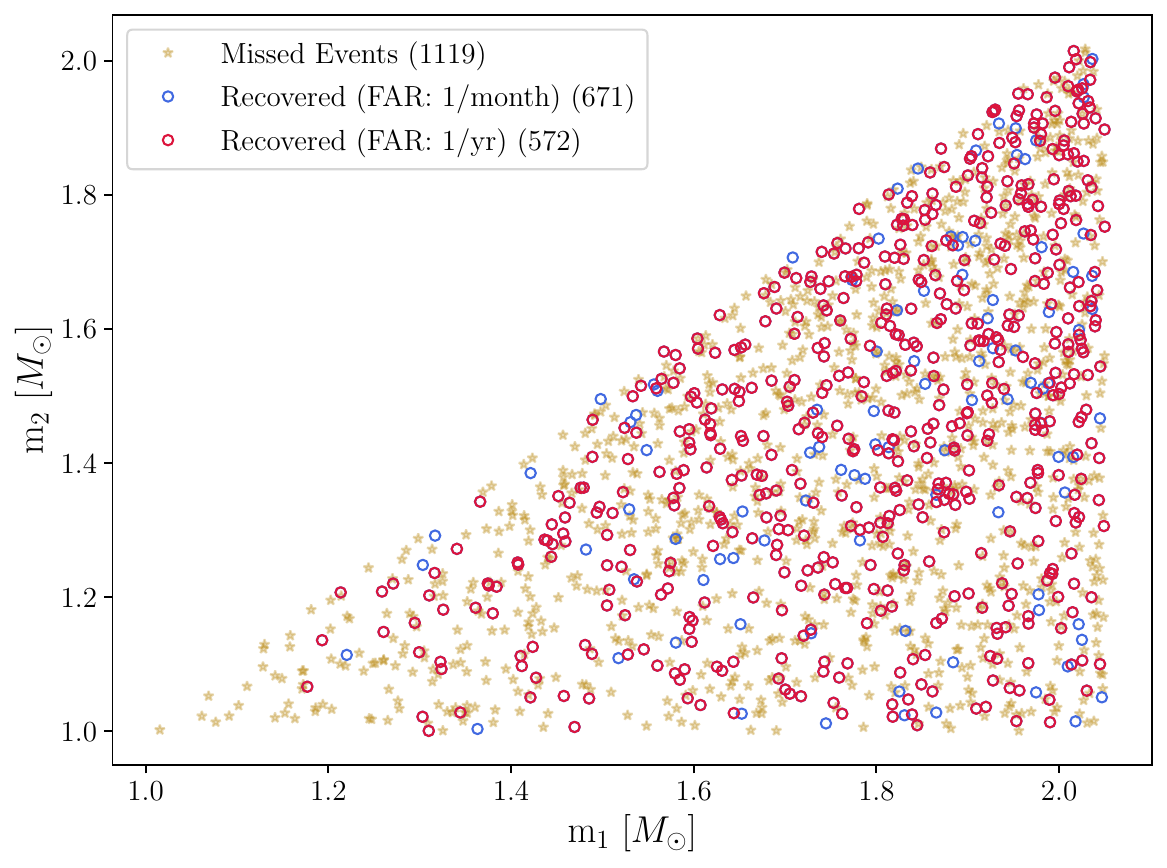}
        \caption{}
        \label{fig:m1-m2}
    \end{subfigure}
    \hfill
    \begin{subfigure}{0.48\textwidth}
        \centering
        \includegraphics[width=\linewidth]{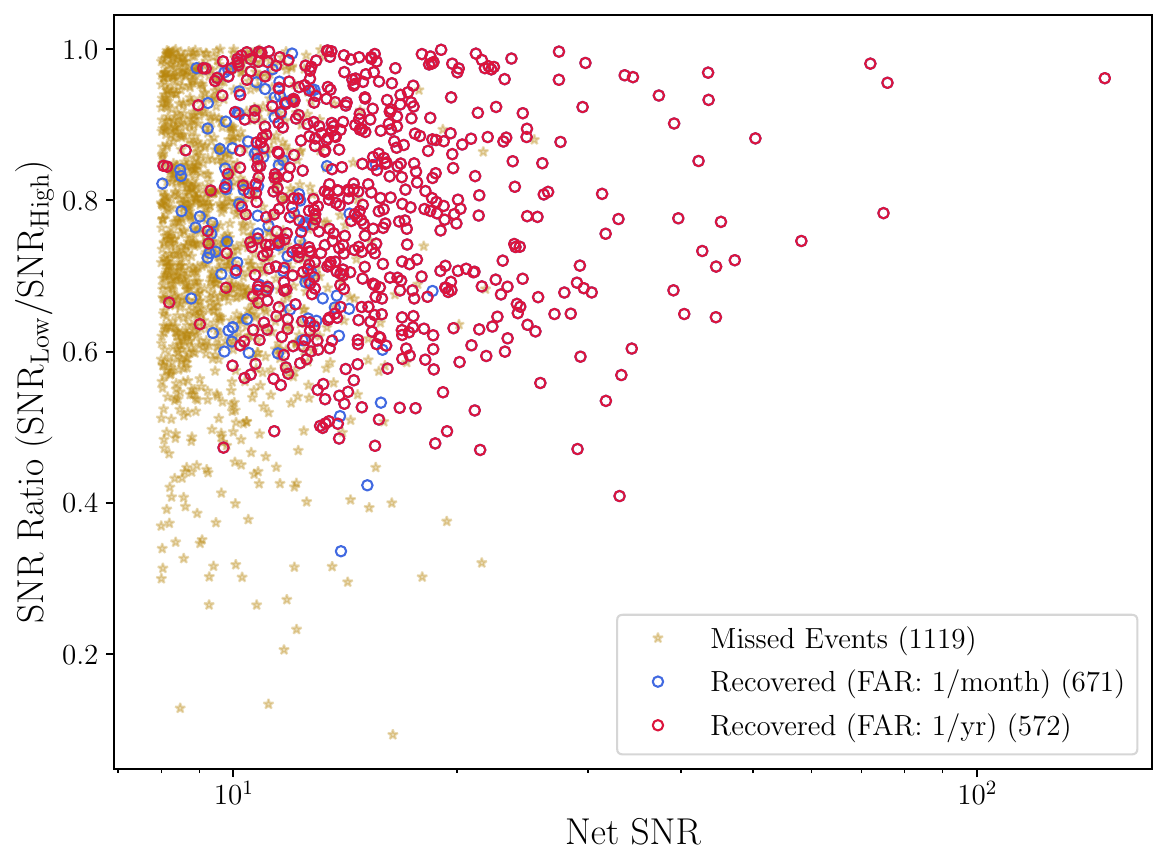}
        \caption{}
        \label{fig:snr-ratio-net-snr}
    \end{subfigure}
    \vspace{0.1cm}
    % Bottom row
    \begin{subfigure}{0.48\textwidth}
        \centering
        \includegraphics[width=\linewidth]{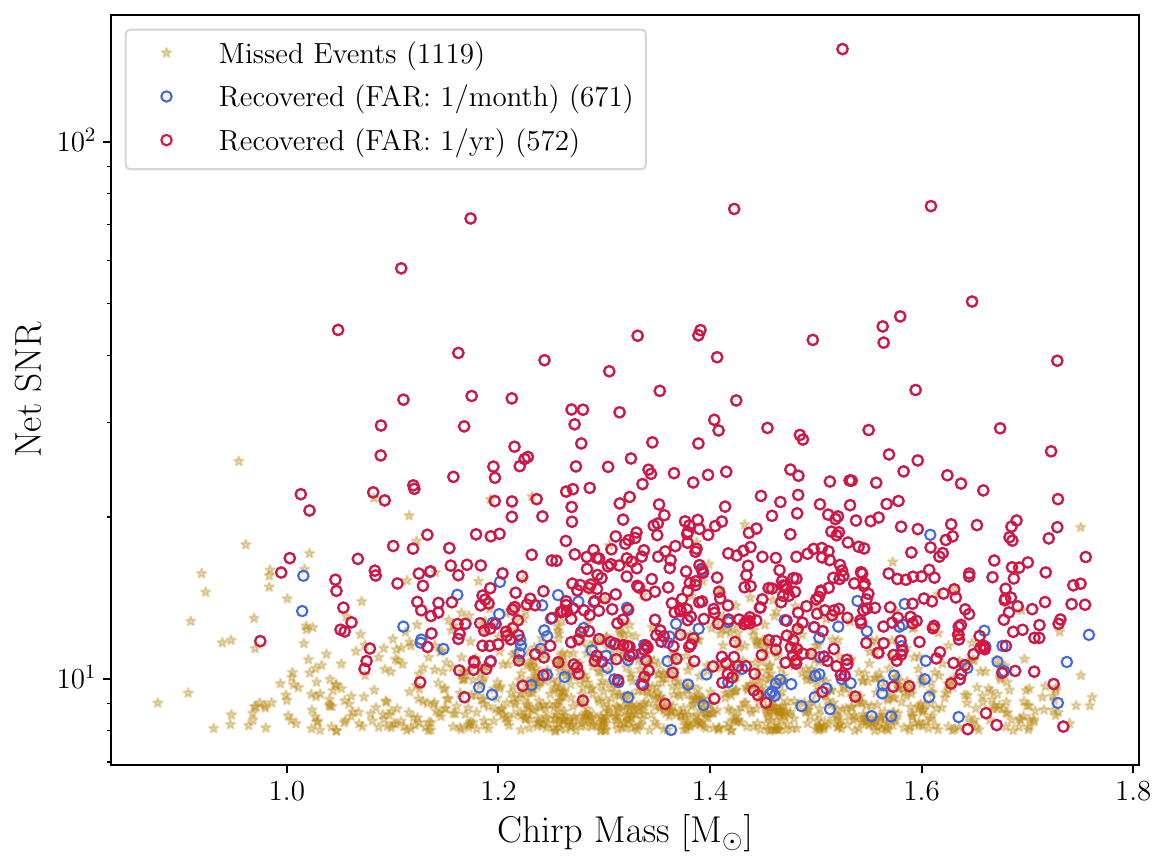}
        \caption{}
        \label{fig:net-snr-mchirp}
    \end{subfigure}
    \hfill
    \begin{subfigure}{0.48\textwidth}
        \centering
        \includegraphics[width=\linewidth]{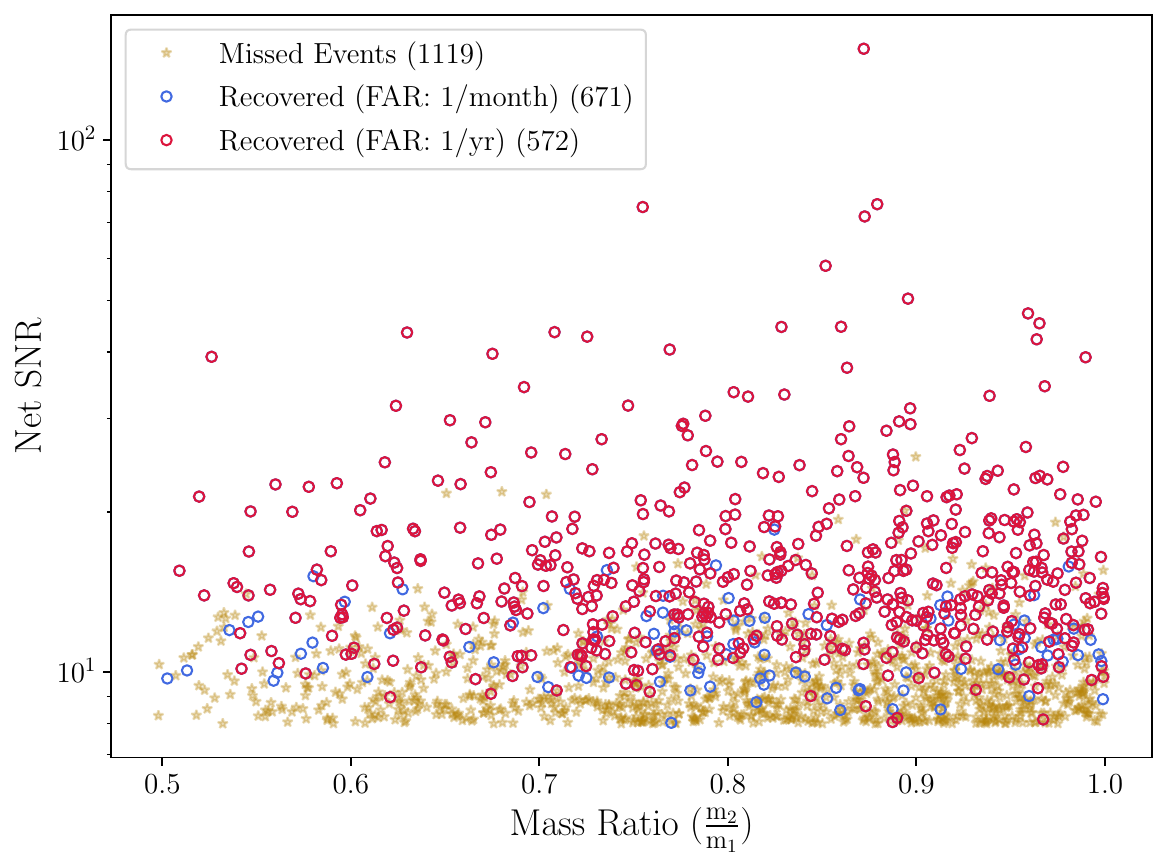}
        \caption{}
        \label{fig:net-snr-mass-ratio}
    \end{subfigure}
    \caption{
    Properties of recovered and missed injections for the binary neutron search over the O3 mock data challenge.
    (a) Distribution of component masses (source frame) for recovered and missed injections. 
    (b) SNR ratio versus network SNR for recovered and missed injections.
    (c) Network SNR as a function of chirp mass (source frame), showing how recovery efficiency varies across the mass parameter space. 
    (d) Network SNR as a function of mass ratio, highlighting regions where injections are preferentially recovered or missed.
    }
    \label{fig:o3-mdc-analysis}
\end{figure*}

The relative performance of {\aframe} further improves for higher-mass systems. For binaries with component masses $m_{1,2} \gtrsim 1.9${\msun}, {\aframe} recovers the largest number of injections at both FAR thresholds, slightly exceeding {\mbta} and {\gstlal} in this regime. This trend is consistent with the one observed in the sensitive volume calculations (See Fig.\,\ref{fig:sensitive-volume}). Higher-mass systems have a considerably shorter signal, making it easier, with the heterodyned representation, to push more signal power toward the merger. At the more strict FAR of $\mathrm{1/year}$, {\aframe} maintains competitive performance relative to the matched-filter pipelines, indicating that the network can preserve sensitivity while suppressing false alarms in realistic detector noise. Overall, these results demonstrate that the heterodyne-based {\aframe} search achieves sensitivity comparable to traditional matched-filter searches while operating with significantly reduced input dimensionality and low-latency inference.

We further characterize {\aframe}'s recovered population in the O3 MDC dataset using the SNR and source parameters as illustrated in Fig.\,\ref{fig:o3-mdc-analysis}. The recovered injections span a broader range of detector SNR ratios, demonstrating that coherent heterodyne channel selection remains effective even when the signal is unevenly distributed across the detector network (Fig.\,\ref{fig:snr-ratio-net-snr}). We observe that the recovered injections cluster toward larger chirp masses and lower network SNRs (Fig.\,\ref{fig:net-snr-mchirp}), indicating that the heterodyned representation more efficiently recovers higher-mass systems. In contrast, lower-chirp-mass systems generally require larger signal power for recovery, consistent with their longer inspiral evolution within the detector frequency band. The recovered events also span the full training mass range in the $m_{1}$--$m_{2}$ plane (Fig.\,\ref{fig:m1-m2}), demonstrating that the network generalizes across the full BNS parameter space rather than biasing towards nearly equal-mass binaries. While the recovery efficiency decreases for highly asymmetric systems (Fig.\,\ref{fig:net-snr-mass-ratio}), the overall dependence on mass ratio remains weak, suggesting that the heterodyned preprocessing is primarily sensitive to the chirp-mass-driven phase evolution of the inspiral. Overall, these trends indicate that the heterodyned representation captures the dominant inspiral structure while remaining robust across detector asymmetries and a broad region of the BNS parameter space.

\section{Suitability for Online Analysis}\label{sec:online_deployment}
\subsection{Offline}
The full training procedure, including data loading, on-the-fly augmentations, dynamic PSD estimation and whitening, and heterodyne preprocessing, requires approximately 96 hours on a single NVIDIA H200 (140 GB) GPU. Training was performed with a batch size of $2000$ and $200$ batches per epoch. The neural-network was optimized for $\sim 2\,{\times}\,10^{5}$ gradient-update steps $(\sim1016\,\,\mathrm{epochs})$, corresponding to $\sim 4\,{\times}\,10^{8}$ training examples processed. Training and validation loss curves are provided in Fig.~\ref{fig:train-val-curves} in Appendix~\ref{appendix:train_val_test}. For offline background estimation over 10-year timeslides mentioned in Section~\ref{sec:performance}, we use Inference-as-a-service (IaaS) using some customizations on top of NVIDIA Triton inference server via the {\tt hermes} library~\cite{hermes}. We deploy the server across two NVIDIA A30 (24 GB) GPUs. Under this setup, the analysis of ten years of background data, together with the testing injection set used to produce Fig.\,\ref{fig:sensitive-volume}, requires approximately 8 days, corresponding to a throughput of $\sim450$\,s of two-detector strain data analyzed per second per GPU. This throughput is comparable to that reported for {\aframe} in the BBH search setting, though the throughput quoted in~\cite{aframe-methods} is on old hardware. The reduction in throughput can be attributed to the additional heterodyne preprocessing stage, the increased number of input channels relative to the two-channel BBH representation, and the longer 8-second analysis windows used in this work. Despite these additional computational requirements, the model remains highly efficient, demonstrating that machine-learning-based BNS searches can be performed while maintaining low inference latency through the use of the optimized neural-network architecture and IaaS.

\subsection{Online}
Once trained, the computational requirements for online deployment are very modest. Real-time inference can be performed on a single NVIDIA A30 GPU (24 GB), operating at an inference sampling rate of 2048\,Hz, which provides sufficient temporal resolution for coalescence-time estimation. The trained model and the related inference buffer utilize $\sim8$ GB of GPU storage. The neural network processes a batch of 128 samples in 22\,ms, corresponding to an average latency of approximately 172\,${\mu\mathrm{s}}$ per sample. The dominant contribution to end-to-end latency arises not from neural-network processing but from the surrounding data-processing steps, including the need to accumulate enough data for whitening and edge cropping, preprocessing, data movement, event identification, and uploading these events to the GraceDB. For {\aframe}, all of these operations contribute $\sim 4.7$ seconds median in total, i.e., from the time GW data arrives at the CIT data center to event candidates registered in GraceDB. In comparison, match filtering CBC pipelines in a production setup take $\sim 10$ seconds for the same steps (see Table 2 of \citet{Chaudhary:2023vec} for O3 MDC performance), though GPU-accelerated variants like SGNL~\cite{sgnl} can be faster.

\section{Conclusion and Outlook}\label{sec:conclusion}
In this work, we have presented a low-latency machine-learning search pipeline for BNS mergers based on the {\aframe} framework. We address the primary challenge for BNS searches -- the long duration of inspiral signals -- by an intuitive preprocessing step that makes the existing {\aframe} search equally suitable for low-mass binaries. We perform a chirp-mass-dependent heterodyne preprocessing step that compresses the coherent inspiral evolution into a compact representation suitable for classification using established neural network architectures such as {\resnet}. This is combined with robust data augmentation and a channel-selection strategy that identifies the most informative heterodyned representations, which enables efficient analysis of long-duration signals using short input windows. We also investigated several alternative preprocessing strategies and input representations; however, the heterodyned representation consistently provided the best performance and computational efficiency. A summary of these exploratory prototypes is provided in Appendix~\ref{appendix:other_rnd}.

We evaluated the sensitivity of our algorithm using the astrophysical metric, sensitive volume, allowing direct comparison with established matched-filter searches from GWTC-3 benchmarks~\cite{Abbott_2023}. Furthermore, we analyzed an independent mock data challenge used by the LVK for production benchmarking~\cite{Chaudhary:2023vec}. While several previous machine-learning-based BNS searches have been developed~\cite{Lin_2019, Krastev_2020, Wei_2021, Krastev_2021, Baltus_2021, Aveiro_2022}, the primary evaluation metric has used ROC curves, which do not directly quantify astrophysical reach. Only one study \cite{schafer2020}, to our knowledge, has reported sensitivity in terms of sensitive distance, and these have generally not achieved sensitivity comparable to that of low-latency matched-filter pipelines used within the LVK. We find that {\aframe} achieves sensitivity comparable to state-of-the-art matched-filter pipelines for low-mass BNS populations and exceeds their sensitivity for higher-mass BNS systems. We validate our findings using two approaches -- the search sensitivities reported in GWTC-3 and the O3 MDC, and reach the same conclusion.  These results demonstrate that machine-learning-based searches can achieve competitive astrophysical performance while maintaining the computational advantages required for low-latency deployment.

In terms of future work, our current implementation remains limited by the duration of the strain segment used for heterodyning. 
%Although the proposed representation successfully compresses a substantial fraction of the inspiral information, 
Thus, extending the heterodyning procedure to longer durations is expected to improve sensitivity to lower-mass systems by recovering
a larger fraction of the inspiral SNR. Subsequent work will explore this direction and also optimize the channel selection procedure.
Furthermore, to produce alert data products, a real-time parameter estimation algorithm, like {\amplfi} for BNS mergers,
enabling rapid sky localization and source property estimation, is needed for end-to-end alerts. Finally,
extending the approach to neutron-star–black-hole (NSBH)
systems, and ultimately enabling a unified all-sky search framework spanning BNS, NSBH, and BBH mergers will be pursued
for production analyses in future observing runs. Such a framework would provide a computationally efficient alternative
to traditional matched-filter searches
while maintaining the sensitivity required for upcoming gravitational-wave observing runs of the LIGO-Virgo-KAGRA detector network as well as for next generation GW instruments.

\begin{acknowledgements}
B.~G. acknowledges support from the ``Accelerated AI Algorithms for Data-Driven Discovery (A3D3)'' Post-Baccalaureate Fellowship at MIT LIGO Lab.
D.~C. would like to thank Javier Roulet for motivating discussions regarding heterodyning during a visit to LIGO-MIT.
The authors thank Carl-Johan Haster for an internal collaboration review of the manuscript. This document is given the LIGO DCC number
P2600320.\footnote{\small\url{https://dcc.ligo.org/LIGO-P2600320}}

The authors acknowledge support from NSF PHY-2117997 (A3D3). 
This work used NCSA-Delta at U. Illinois through allocation PHY-240078 from the Advanced Cyberinfrastructure Coordination Ecosystem: Services \& Support (ACCESS) program, which is supported by National Science Foundation grants \#2138259, \#2138286, \#2138307, \#2137603, and \#2138296. The authors are also grateful for computational resources provided by LIGO Laboratory supported by NSF grants PHY-0757058 and PHY-0823459. This research has made use of data or software obtained from the Gravitational Wave Open Science Center~\footnote{\url{https://gwosc.org/}}~\cite{gwosc2, gwosc3}, a service of the LIGO Scientific Collaboration, the Virgo Collaboration, and KAGRA. This material is based upon work supported by NSF's LIGO Laboratory which is a major facility fully funded by the NSF, as well as the Science and Technology Facilities Council (STFC) of the United Kingdom, the Max-Planck-Society (MPS), and the State of Niedersachsen/Germany for support of the construction of Advanced LIGO and construction and operation of the GEO600 detector. Additional support for Advanced LIGO was provided by the Australian Research Council. Virgo is funded, through the European Gravitational Observatory (EGO), by the French Centre National de Recherche Scientifique (CNRS), the Italian Instituto Nazionale di Fisica Nucleare (INFN) and the Dutch Nikhef, with contributions by institutions from Belgium, Germany, Greece, Hungary, Ireland, Japan, Monaco, Poland, Portugal, Spain. KAGRA is supported by Ministry of Education, Culture, Sports, Science and Technology (MEXT), Japan Society for the Promotion of Science (JSPS) in Japan; National Research Foundation (NRF) and Ministry of Science and ICT (MSIT) in Korea; Academia Sinica (AS) and National Science and Technology Council (NSTC) in Taiwan.

\end{acknowledgements}

\appendix
\section{Training, Validation, Testing}\label{appendix:train_val_test}
{\aframe} is trained on real detector strain rather than stationary Gaussian noise, ensuring that training, validation, and testing reflect realistic detector conditions. Though this work used public observatory data from GWOSC, for an observing run non-public data that is available
during an engineering run is used to generate background. Only data segments that satisfy analysis-ready data-quality requirements and are coincident between the H1 and L1 detectors are retained. The dataset is partitioned with disjoint training and testing segments. This separation mimics the constraints of real-time deployment, requiring the model to generalize to future data without prior exposure. The pipeline operates on a two-detector network, jointly analyzing strain data from the H1 and L1 interferometers.
These data streams are provided simultaneously as combined input channels, enabling the model to learn coherent and coincident features
across the detector network. This enables the network to learn consistent features across astrophysical signals, such as correlated
phase evolution, while distinguishing them from uncorrelated noise transients, thereby improving robustness to detector-specific noise.
The Virgo detector, V1, due to its lower range does not contribute significantly to the \emph{detection} of an event. It is, however,
important for parameter estimation.

During training, a training batch is divided into foreground and background data, determined
randomly and dynamically at every iteration. For injected signals in the foreground, {\aframe} has the ability to
dynamically generate the full GW waveform on-the-fly from a motivated astrophysical prior. Alternatively, a precomputed
waveform bank based on the CBC intrinsic parameters, like mass and spin, can be supplied. While the former
allows a more diverse training data with full waveforms generated on-the-fly during training, the latter allows
for incorporating expensive waveforms with more physics that are not a part of the {\mlgw} framework.
Irrespective of how the waveform polarizations are generated, they are then projected
based on the detectors based on extrinsic parameters like sky-coordinates, distance, polarization angle dynamically,
which allows each intrinsic waveform to appear under a wide range of observational conditions, increasing the variability
of the training data.  During validation and testing, rejection sampling is employed to preferentially retain signals
above a minimum signal-to-noise ratio (SNR) threshold, called ``hopeless'' threshold.
This reduces statistical uncertainty in sensitivity estimates by focusing the evaluation on signals
that contribute meaningfully to detection performance.

\begin{figure}[h]
    \centering
    \includegraphics[width=0.9\textwidth]{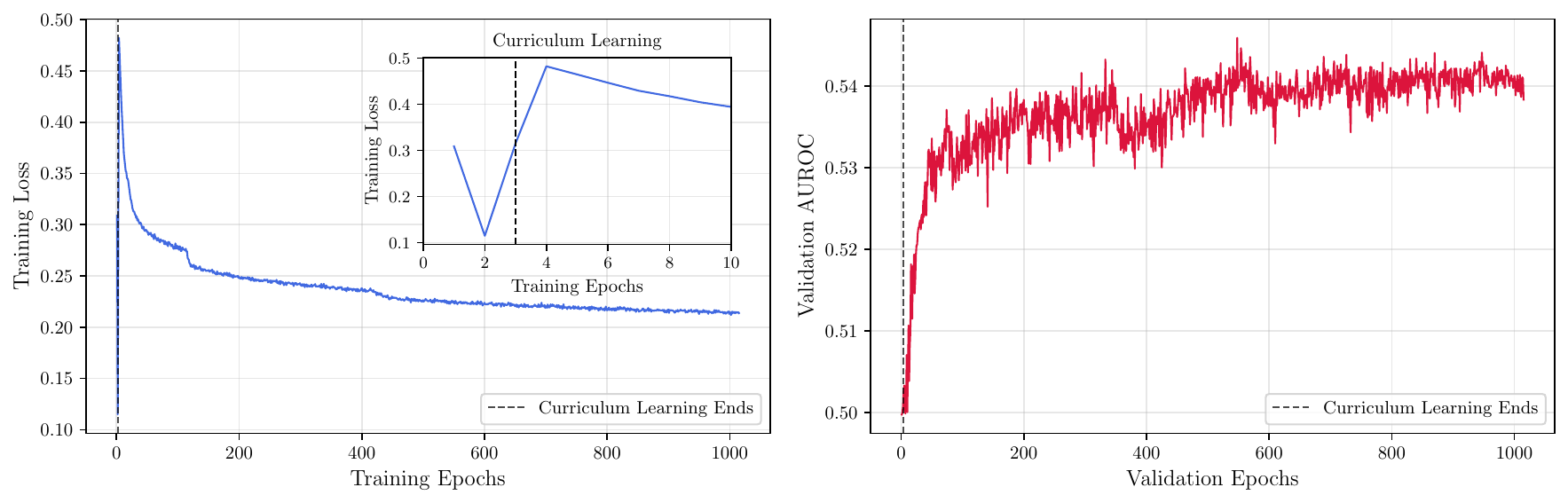}
    \caption{Training loss (left) and validation AUROC (right) as a function of training epoch. The network was trained using a curriculum-learning strategy in which injected signals were initially restricted to louder events, with network SNRs drawn from a power-law distribution between 30 and 100 during the first three epochs, before transitioning to the full training distribution with network SNRs drawn from a power-law distribution between 8 and 100. The rapid decrease in training loss during the curriculum phase, followed by a temporary increase after the transition, reflects the introduction of more challenging lower-SNR signals into the training set. After this transition, the training loss continues to decrease while the validation AUROC remains stable, indicating successful convergence of the model.}
    \label{fig:train-val-curves}
\end{figure}

The training involves a binary classification task where the neural network learns to classify
segments that are tagged positively or negatively. The loss function that is used is the binary-cross
entropy. The loss curve for a typical training is shown in Fig.~\ref{fig:train-val-curves}. The first
few epochs follow an easy curriculum where the model is preferentially shown louder injections that
are easier to classify. Following which the training distribution is annealed to one that more
representative of real observations. We use the area under the ROC curve as a validation metric.
This is a standard for classification tasks. We note that the absolute value of the area under
the curve (y-axis in right panel of Fig.~\ref{fig:train-val-curves}) not approaching unity
is an expected behavior as this efficiency value is calculated at fixed false alarm of $10^{-3}$.
This is significantly higher compared to usual operations false alarms of GW detectors. In our
case, the feature of interest is the increasing value of this metric with training epochs as the
network learns to better classify signals and background. We also do not use the area under the
curve as a way to compare against established search algorithms. Instead, our testing step involves
calculating a sensitive volume in a monte carlo sense, which is approach taken by match-filter
algorithms to assess their performance. Also note that training and validation emph{does not}
involve streaming the data, which is only done for inference.

During inference, the online deployment setup involves operating on continuous stream of detector data
using a sliding-window approach, evaluating the neural network on overlapping segments to produce a
timeseries of detection scores. These scores serve as a ranking statistic analogous to those used in
traditional matched-filter searches. Candidate events are identified by clustering peaks in the network
output and assigning statistical significance by comparing them to a background distribution estimated
from timeslides (See Section IX in \cite{aframe-methods} and Section \ref{sec:performance} above).
This framework enables the assignment of false alarm rates (FARs) to candidate detections in a manner
consistent with established gravitational-wave search pipelines. A main advantage of this approach is
its computational efficiency. Once trained, the neural network can process data at or above real-time
rates with relatively modest hardware requirements (one NVIDIA 24\,GB A30 GPU) and does not require
frequent retraining (longevity section of aframe paper), thereby preserving the model's longevity.
This makes the pipeline well-suited for low-latency applications, where rapid identification of GW
events is critical for enabling multi-messenger follow-up observations.

\section{Alternative Methods Explored}\label{appendix:other_rnd}
\subsection{Decimator/Multi-Banding}
We explored a multi-banding strategy motivated by the time-frequency evolution exhibited by BNS mergers. These binaries remain in the detector band for $\mathcal{O}(\mathrm{mins.})$, with the inspiral phase occupying a large fraction of the signal duration. During inspiral, the signal evolves slowly in time while sweeping upward in frequency, implying that different segments of the signal contain information at different characteristic time and frequency scales. As a result, the information content of the signal is not distributed uniformly across the observation window.

To exploit this structure, we segmented each 20\,s analysis window into two components. The final 4\,s including the merger (placed randomly in a 1\,s segment from the right edge of the analysis window), which contain the highest-frequency and highest-SNR portion of the signal, were retained at 2048\,Hz and provided directly to a neural network as a timeseries input. The preceding 16\,s, dominated by lower-frequency inspiral content, were downsampled to 512\,Hz and converted into a spectrogram representation. The motivation was that a time-frequency representation would more naturally capture the chirping morphology of inspiral signals while suppressing noise features that lack coherent frequency evolution. We therefore trained a multi-branch architecture, with one network processing the 4\,s timeseries segment and a second network processing the 16\,s spectrogram segment, each trained using binary cross-entropy (BCE) loss. The scores from the two neural networks were combined using a simple average.

During training and testing, we saw that the 16\,s inspiral branch contributed no additional discriminative power beyond what was already provided by the final 4\,s segment. We also found that the network relied predominantly on the louder merger-adjacent portion of the signal, suggesting that the lower-frequency inspiral information was not being effectively utilized by the model. The multi-band prototype for the BNS search algorithm did not yield measurable sensitivity improvements over the model trained only on the final 4\,s of strain data (See Fig.~\ref{fig:bns-prototype-sv}).

We adopted a simpler configuration based on a 4\,s analysis window. Although the 4\,s configuration yielded the best performance, it remained suboptimal based on the sensitive volume testing metric. We think this is because a significant fraction of the SNR accumulates during the earlier inspiral phase. More broadly, these experiments highlighted a limitation of extending the analysis window within our neural-network framework. As the window length increases, the amount of background information grows substantially, and the network appears to focus primarily on the most prominent features near the merger rather than learning the time-frequency evolution of the signal. Capturing information distributed across tens of seconds may therefore require architectures specifically designed for long-range temporal dependencies or explicit time-frequency reasoning, rather than conventional {\resnet} architectures operating on extended strain segments.

Motivated by the hypothesis that the observed limitations were partially architectural rather than representational, we investigated Temporal Convolutional Networks ({\tcn}) as an alternative to {\resnet} for the 4\,s timeseries input. {\tcn}s employ dilated causal convolutions that are well-suited for modeling sequential data while retaining many of the computational advantages of convolutional architectures. We explored a few network configurations and hyperparameter choices; however, the resulting models did not achieve performance comparable to the {\resnet} and exhibited less stable training behavior. Consequently, further development of the {\tcn}-based approach was not pursued.

\subsection{Long Window Spectrogram View}
As discussed earlier, BNS mergers exhibit a chirp-like evolution in time-frequency space, with the signal frequency increasing as the binary inspirals toward merger. Motivated by this behavior, we explored a spectrogram-based representation that could expose the signal's time-frequency structure to the network. For this study, we used an 8\,s analysis window in which the merger time was randomly placed within the final 1\,s of the segment. The strain timeseries was converted into a two-dimensional spectrogram spanning both time and frequency, and a 2D {\resnet} architecture ({\resnettwod}) was trained as an image-classification model using the BCE loss.

\begin{figure}[h]
    \centering
    \includegraphics[width=0.5\textwidth]{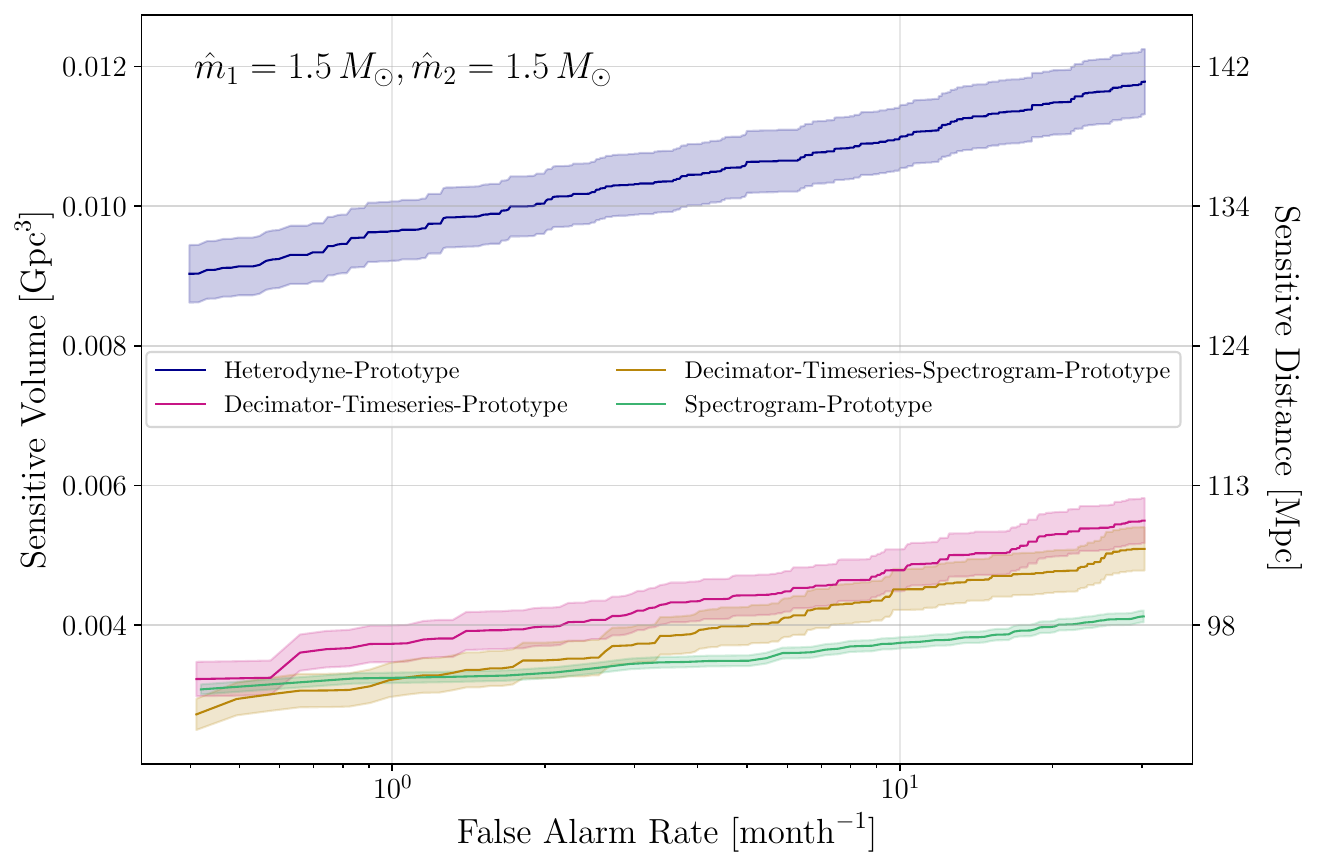}
    \caption{Sensitive volume as a function of false alarm rates for the different exploratory approaches designed to incorporate longer-duration inspiral information for binary neutron stars.}
    \label{fig:bns-prototype-sv}
\end{figure}

The primary motivation for this approach was that GW signals occupy a coherent trajectory in time-frequency space, whereas detector noise lacks a similarly structured evolution. While this approach showed promising qualitative behavior and successfully learned signal-like features, it did not achieve the same sensitivity as the baseline 4\,s timeseries model (See Fig.~\ref{fig:bns-prototype-sv}). We speculate that this performance gap arises from a combination of information loss, introduced by the finite time-frequency resolution of the spectrogram representation, and the increased difficulty of extracting weak inspiral features over a longer analysis window containing substantially more background noise. Although the spectrogram representation provided a more physically interpretable view of the signal evolution, it did not offer a measurable improvement in detection performance and was not pursued further.

Altogether, these experiments suggest that simply increasing the analysis duration or introducing explicit time-frequency representations is insufficient to improve sensitivity within the architectures and data representations explored above. This indicates that future improvements may require either (1) architectures capable of efficiently capturing the long-duration evolution of BNS signals while remaining robust to the increased background complexity associated with longer analysis windows, or (2) more effective preprocessing and feature-representation techniques that transform the signal into a form that is more readily distinguishable from background noise. Exploring these directions remains an important avenue for extending machine-learning searches to longer-duration gravitational-wave signals.

\bibliography{references}

\end{document}